\RequirePackage{lineno}
\documentclass[a4paper,11pt,epsfig]{article}

\usepackage{jheppub} 
                     
\usepackage[T1]{fontenc} 
\usepackage{graphicx,color,dcolumn,booktabs,bm}
\usepackage{longtable,lscape}
\usepackage{amsmath}
\usepackage{indentfirst}
\usepackage{epsfig}
\usepackage{epstopdf}   
\usepackage{slashed}  
\usepackage{color}
\usepackage{multirow}
\usepackage{soul}  
\usepackage{graphicx,color,dcolumn,booktabs,bm}
\usepackage{verbatim}
\usepackage{orcidlink}
\usepackage{appendix}

\newcommand{\BR}{{\mathcal B}}

\newcommand{\xicz}{\Xi_{c}^{0}}
\newcommand{\xiz}{\Xi^{0}}
\newcommand{\hz}{h^{0}}
\newcommand{\piz}{\pi^{0}}
\newcommand{\etap}{\eta^{\prime}}
\newcommand{\xizhz}{\xiz\hz}
\newcommand{\xizpiz}{\xiz\piz}
\newcommand{\xizeta}{\xiz\eta}
\newcommand{\xizetap}{\xiz\etap}
\newcommand{\xim}{\Xi^{-}}
\newcommand{\pim}{\pi^{-}}
\newcommand{\pip}{\pi^{+}}
\newcommand{\xipi}{\xim\pip}
\newcommand{\etal}{{\em et al.}}
\newcommand{\ep}{e^{+}}
\newcommand{\e}{e^{-}}
\newcommand{\ee}{\ep\e}
\newcommand{\eetocc}{\ee\to c \bar{c}}
\newcommand{\mevcs}{{\rm MeV}/c^2}
\newcommand{\dig}{\gamma\gamma}

\title{\boldmath Measurements of the branching fractions of $\Xi_{c}^{0}\to\Xi^{0}\pi^{0}$, $\Xi_{c}^{0}\to\Xi^{0}\eta$, and $\Xi_{c}^{0}\to\Xi^{0}\eta^{\prime}$ and asymmetry parameter of $\Xi_{c}^{0}\to\Xi^{0}\pi^{0}$ }

\preprint{\vbox{ \hbox{   }
\hbox{Belle II Preprint 2024-015 }
\hbox{KEK Preprint 2024-9}
}}

\collaboration{The Belle and Belle II Collaborations}
  \author{I.~Adachi\,\orcidlink{0000-0003-2287-0173},} 
  \author{L.~Aggarwal\,\orcidlink{0000-0002-0909-7537},} 
  \author{H.~Aihara\,\orcidlink{0000-0002-1907-5964},} 
  \author{N.~Akopov\,\orcidlink{0000-0002-4425-2096},} 
  \author{A.~Aloisio\,\orcidlink{0000-0002-3883-6693},} 
  \author{N.~Althubiti\,\orcidlink{0000-0003-1513-0409},} 
  \author{N.~Anh~Ky\,\orcidlink{0000-0003-0471-197X},} 
  \author{D.~M.~Asner\,\orcidlink{0000-0002-1586-5790},} 
  \author{H.~Atmacan\,\orcidlink{0000-0003-2435-501X},} 
  \author{T.~Aushev\,\orcidlink{0000-0002-6347-7055},} 
% ...

\abstract{We present a study of $\Xi_{c}^{0}\to\Xi^{0}\pi^{0}$, $\Xi_{c}^{0}\to\Xi^{0}\eta$, and $\Xi_{c}^{0}\to\Xi^{0}\eta^{\prime}$ decays using the Belle and Belle~II data samples, which have integrated luminosities of 980~$\mathrm{fb}^{-1}$ and 426~$\mathrm{fb}^{-1}$, respectively.
We measure the following relative branching fractions
$${\mathcal B}(\Xi_{c}^{0}\to\Xi^{0}\pi^{0})/\BR(\Xi_{c}^{0}\to\Xi^{-}\pi^{+}) = 0.48 \pm 0.02 ({\rm stat}) \pm 0.03 ({\rm syst}) ,$$
$${\mathcal B}(\Xi_{c}^{0}\to\Xi^{0}\eta)/\BR(\Xi_{c}^{0}\to\Xi^{-}\pi^{+}) = 0.11 \pm 0.01 ({\rm stat}) \pm 0.01 ({\rm syst}) ,$$
$${\mathcal B}(\Xi_{c}^{0}\to\Xi^{0}\eta^{\prime})/\BR(\Xi_{c}^{0}\to\Xi^{-}\pi^{+}) = 0.08 \pm 0.02 ({\rm stat}) \pm 0.01 ({\rm syst}) $$
for the first time, where the uncertainties are statistical ($\rm stat$) and systematic ($\rm syst$).
By multiplying by the branching fraction of the normalization mode, ${\mathcal B}(\Xi_{c}^{0}\to\Xi^{-}\pi^{+})$, we obtain the following absolute branching fraction results 
$$\BR(\xicz\to\xizpiz) =  (6.9 \pm 0.3 ({\rm stat}) \pm 0.5 ({\rm syst}) \pm 1.5 ({\rm norm})) \times 10^{-3},$$
$$\BR(\xicz\to\xizeta) =  (1.6 \pm 0.2 ({\rm stat}) \pm 0.2 ({\rm syst}) \pm 0.4 ({\rm norm})) \times 10^{-3},$$
$$\BR(\xicz\to\xizetap) = (1.2 \pm 0.3 ({\rm stat}) \pm 0.1 ({\rm syst}) \pm 0.3 ({\rm norm})) \times 10^{-3},$$
where the third uncertainties are from $\BR(\xicz\to\xipi)$.
The asymmetry parameter for $\Xi_{c}^{0}\to\Xi^{0}\pi^{0}$ is measured to be $\alpha(\Xi_{c}^{0}\to\Xi^{0}\pi^{0}) = -0.90\pm0.15({\rm stat})\pm0.23({\rm syst})$.
}


\begin{document}
\maketitle
\flushbottom

\section{Introduction}
Charmed baryons provide an interesting dynamical system to study the interplay of strong and weak interactions.
Recently, there have been several impactful measurements for the $\xicz$ baryon.
In particular, 
the absolute branching fractions of several $\xicz$ decay modes, especially the normalization mode $\xicz\to\xipi$, have been measured~\cite{xic0absbf2019}, 
allowing for the determination of branching fractions for other channels using ratios of branching fractions. 
In addition, the Belle experiment has recently measured branching fractions and decay asymmetry parameters for several Cabibbo-favored (CF) decays, including the two-body $\xicz\to B~V$ decays $\xicz\to \Lambda {\bar K}^{*0}$, $\Sigma^0 {\bar K}^{*0}$, and $\Sigma^+ K^{*-}$~\cite{xic02hypKstar2021} as well as the branching fractions for the two-body $\xicz\to B~P$ decays $\xicz\to \Lambda K^0_S$, $\Sigma^0 K^0_S$, and $\Sigma^+ K^-$~\cite{xic02hypK2022}, where $B$, $V$, and $P$ represent light baryons, vector mesons, and pseudoscalar mesons, respectively.
Additional measurements of $\xicz$ branching fractions and decay asymmetry parameters may allow for a more complete description of the dynamics of the strong and weak interactions in charmed baryon decays.

\begin{figure}[htbp]
	\centering

	\includegraphics[width=7cm]{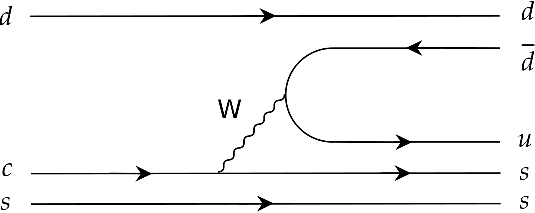}\put(-150,90){\bf (a)}	\hspace{1cm}
	\includegraphics[width=7cm]{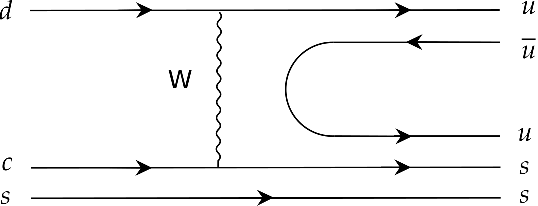}\put(-150,90){\bf (b)}
	\caption{Feynman diagrams for (a) internal $W$-emission and (b) $W$-exchange in $\xicz\to\xizhz$ decays~\cite{charmedBaryon2022}.}
	\label{feynman}
\end{figure}

In hadronic weak-interaction decays of charmed baryons, nonfactorizable amplitudes arising from internal $W$-emission and $W$-exchange quark-level processes play an essential role and lead to difficulties for theoretical predictions~\cite{charmedBaryon2022}.
Figure~\ref{feynman} shows the Feynman diagrams for the internal $W$-emission and $W$-exchange amplitudes in CF $\xicz\to\xizhz$ decays, to which only the nonfactorizable amplitudes contribute~\cite{charmedBaryon2022}.
In the following, $\hz$ refers to $\piz$, $\eta$, or $\etap$ mesons.
Various approaches have been developed to describe the nonfactorizable effects, including 
the covariant confined quark model~\cite{theory1quark1992,theory5quark1998}, 
the pole model~\cite{theory2pole1992,theory3poleca1993,theory4poleca1994,theory10poleca2020}, 
current algebra (CA)~\cite{theory3poleca1993,theory4poleca1994,theory6ca1999,theory10poleca2020}, 
and $\rm SU(3)_F$ flavor symmetry~\cite{theory7su3f2018,theory8su3f2019,theory9su3f2020,theory11su3f2022,theory12su3f2022,theory13su3f2023,theory14su3f2023,theory15su3f2024,theory16su3f2024} based treatments.
Theoretical predictions for the branching fractions of $\xicz\to\xizhz$ decays based on these approaches are listed in table~\ref{theoryPredict}.
Measurements of the branching fractions for $\xicz\to\xizhz$ decays will help to clarify the theoretical picture.

\begin{table}[htbp!]
	\caption{Theoretical predictions for the branching fractions and decay asymmetry parameters for $\xicz\to\xizhz$ decays. Branching fractions are given in units of $10^{-3}$. }
	\vspace{0.2cm}
	\label{theoryPredict}
	\centering
	\footnotesize
	\begin{tabular}{l c c c c c}
		\hline\hline
		Reference & Model & $\BR(\xicz\to\xizpiz)$ & $\BR(\xicz\to\xizeta)$ & $\BR(\xicz\to\xizetap)$ & $\alpha(\xicz\to\xizpiz)$  \\
		\hline
		K$\rm\ddot{o}$rner, Kr$\rm\ddot{a}$mer~\cite{theory1quark1992}	& Quark 						& 0.5						& 3.2						& 11.6							& 0.92						 \\
		Ivanov~\etal~\cite{theory5quark1998} 							& Quark							& 0.5						& 3.7						& 4.1							& 0.94						 \\
		Xu, Kamal~\cite{theory2pole1992} 								& Pole							& 7.7						& -							& -								& 0.92						\\
		Cheng, Tseng~\cite{theory3poleca1993}  							& Pole							& 3.8						& -							& -								& $-$0.78						\\
		{\.Z}enczykowski~\cite{theory4poleca1994} 						& Pole							& 6.9						& 0.1						& 0.9 							& 0.21						\\
		Zou~\etal~\cite{theory10poleca2020} 							& Pole							& 18.2						& 26.7						& -								& $-$0.77 							\\
		Sharma, Verma~\cite{theory6ca1999} 								& CA							& -							& -							& -								& $-$0.8						\\
		Cheng, Tseng~\cite{theory3poleca1993}  							& CA							& 17.1						& -							& -								& 0.54						\\
		Geng~\etal~\cite{theory7su3f2018} 								& $\rm SU(3)_F$					& $4.3\pm0.9$				& $1.7^{+1.0}_{-1.7}$		& $8.6^{+11.0}_{-6.3}$			& -							\\
		Geng~\etal~\cite{theory8su3f2019} 								& $\rm SU(3)_F$					& $7.6\pm1.0$				& $10.3\pm2.0$ 				& $9.1\pm4.1$					& $-1.00^{+0.07}_{-0.00}$	 \\
		Zhao~\etal~\cite{theory9su3f2020} 								& $\rm SU(3)_F$					& $4.7\pm0.9$ 				& $8.3\pm2.3$				& $7.2\pm1.9$					& - 						\\
		Huang~\etal~\cite{theory11su3f2022} 							& $\rm SU(3)_F$					& $2.56\pm0.93$				& -							& -								& $-0.23\pm0.60$			\\
		Hsiao~\etal~\cite{theory12su3f2022}  							& $\rm SU(3)_F$ 				& $6.0\pm1.2$				& $4.2^{+1.6}_{-1.3}$		& -								& -							\\
		Hsiao~\etal~\cite{theory12su3f2022}  							& $\rm SU(3)_F$-breaking 		& $3.6\pm1.2$ 				& $7.3\pm3.2$				& -								& -							\\
		Zhong~\etal~\cite{theory13su3f2023}  							& $\rm SU(3)_F$					& $1.13^{+0.59}_{-0.49}$	& $1.56\pm1.92$ 			& $0.683^{+3.272}_{-3.268}$		& $0.50^{+0.37}_{-0.35}$		\\
		Zhong~\etal~\cite{theory13su3f2023} 							& $\rm SU(3)_F$-breaking		& $7.74^{+2.52}_{-2.32}$	& $2.43^{+2.79}_{-2.90}$	& $1.63^{+5.09}_{-5.14}$		& $-0.29^{+0.20}_{-0.17}$	  \\
		Xing~\etal~\cite{theory14su3f2023}								& $\rm SU(3)_F$					& $1.30\pm0.51$				& -							& -								& $-0.28\pm0.18$ 			\\
		Geng~\etal~\cite{theory15su3f2024}								& $\rm SU(3)_F$					& $7.10\pm0.41$				& $2.94\pm0.97$				& $5.66\pm0.93$					& $-0.49\pm0.09$			\\
		Zhong~\etal~\cite{theory16su3f2024}								& Diagrammatic-$\rm SU(3)_F$	& $7.45\pm0.64$				& $2.87\pm0.66$				& $5.31\pm1.33$					& $-0.51\pm0.08$			\\
		Zhong~\etal~\cite{theory16su3f2024}								& Irreducible-$\rm SU(3)_F$		& $7.72\pm0.65$				& $2.28\pm0.53$				& $5.66\pm1.62$					& $-0.51\pm0.09$			\\
		\hline\hline
	\end{tabular}
\end{table}

In addition to the branching fraction measurement, parity violation can also be studied.
In weak-interaction decays, the interference between the parity-violating and parity-conserving amplitudes leads to an asymmetry in the angular decay distribution, which can be quantified by the parameter $\alpha$.
In $\xicz\to\xizhz$ decays, $\alpha$ can be extracted by fitting to the $\xicz$ decay angular distribution, using the differential decay rate function~\cite{alphaFunction}, 
\begin{equation}\label{eq:1}
\frac{dN}{d\cos\theta_{\xiz}}\propto 1+ \alpha(\xicz\to\xizhz)\alpha(\xiz\to\Lambda\piz)\cos\theta_{\xiz},
\end{equation}
where $\alpha(\xiz\to\Lambda\piz)$ is the asymmetry parameter for $\xiz\to\Lambda\piz$ and $\cos\theta_{\xiz}$ is the angle between the $\Lambda$ momentum vector and the direction opposite to the $\xicz$ momentum vector in the $\xiz$ rest frame.
Predictions for $\alpha(\xicz\to\xizpiz)$ from various models are also listed in table~\ref{theoryPredict}. 
In addition to $\alpha(\xicz\to B~V)$~\cite{xic02hypKstar2021}, $\alpha(\xicz\to\xipi)$ has also been measured by CLEO and Belle~\cite{xic02xipiAlpha2001,xic0semilptANDalpha2021}.

In this paper, we present the first measurement of the branching fractions for $\xicz\to\xizpiz$, $\xicz\to\xizeta$, and $\xicz\to\xizetap$ decays, and the asymmetry parameter of the $\xicz\to\xizpiz$ decay.
The $\xicz\to\xipi$ decay is taken as the normalization mode for absolute branching fraction measurements. 
The signal yields used for branching fraction measurements are extracted from fits to the invariant mass distributions of fully reconstructed $\xicz$ candidates.
The asymmetry parameter $\alpha(\xicz\to\xizpiz)$ is obtained from a linear fit to the $\xicz$ signal yield as a function of $\cos\theta_{\Xi^0}$. 
This analysis combines data samples with integrated luminosities of 980~$\mathrm{fb}^{-1}$ and 426~$\mathrm{fb}^{-1}$ collected with the Belle and Belle~II detectors operating at the KEKB and SuperKEKB asymmetric-energy $\ee$ colliders, respectively.
Charge-conjugate modes are implied throughout the paper.

\section{Belle and Belle~II detectors}

The Belle detector~\cite{Belle1,Belle2} operated from 1999 to 2010 at the KEKB asymmetric-energy $\ee$ collider~\cite{KEKB1,KEKB2}.
Belle was a large cylindrical solid-angle magnetic spectrometer that consisted of a silicon vertex detector, a central drift chamber, an array of aerogel threshold Cherenkov counters, a barrel-like arrangement of time-of-flight scintillation counters, an electromagnetic calorimeter (ECL) comprised of CsI(Tl) crystals located inside a superconducting solenoid coil that provided a $1.5~\hbox{T}$ axial magnetic field, and an iron flux return placed outside the coil, instrumented with resistive-plate chambers to detect $K^{0}_{L}$ mesons and to identify muons.
A detailed description of the detector can be found in refs.~\cite{Belle1,Belle2}.

The Belle~II detector~\cite{BelleII} is located at the interaction point of the SuperKEKB asymmetric-energy $\ee$ collider~\cite{superKEKB}.
Belle~II is an upgraded version of the Belle detector and consists of several new subsystems and substantial upgrades to others.
The new vertex detector includes two inner layers of
pixel sensors and four outer layers of double-sided silicon microstrip sensors.
For the data sample used in this analysis, the second pixel layer was incomplete, covering only one sixth of the azimuthal angle.
A new central drift chamber surrounding the vertex detector is used to measure the momenta and electric charges of charged particles. 
A time-of-propagation detector in the barrel and an aerogel ring-imaging Cherenkov detector in the forward endcap provide information for the identification of charged particles, supplemented by ionization energy loss measurements in the central drift chamber.
To cope with the higher beam-induced background environment at Belle II, the ECL readout electronics has been upgraded. 
The superconducting solenoid coil and the iron flux return for Belle are reused in Belle II, 
with some of the resistive-plate chambers in the $K^{0}_{L}$ and muon detector replaced by plastic scintillator modules.

The $z$ axis of the cylindrical laboratory frame is defined as the central solenoid axis with the positive direction toward the $e^-$ beam, common to Belle and Belle II.

\section{Data sample \label{datasets}}
This measurement uses data recorded at center-of-mass (c.m.) energies at or near the $\Upsilon(1S)$, $\Upsilon(2S)$, $\Upsilon(3S)$, $\Upsilon(4S)$, and $\Upsilon(5S)$ resonances by the Belle detector, and at or near the $\Upsilon(4S)$ and at 10.75~GeV by the Belle II detector.
The data samples correspond to integrated luminosities of 980~$\mathrm{fb}^{-1}$ and 426~$\mathrm{fb}^{-1}$ for Belle and Belle~II, respectively.

Monte Carlo (MC) samples of simulated events are used to optimize signal selection criteria, calculate the reconstruction efficiency, and investigate possible background sources.
Signal events are generated using the {\sc pythia}~\cite{pythia1,pythia2} and {\sc evtgen}~\cite{evtgen} software packages via $\eetocc$,
where one of the charm quarks is required to hadronize into a $\xicz$ baryon.
Simulated $\xicz\to\xizhz/\xipi$ decays are generated with a phase space model. 
To determine the efficiency for the branching fraction measurement, the simulated signal samples are weighted according to eq.~(\ref{eq:1}) and the measured values of $\alpha$.
Due to the small sample size, $\alpha$ is not measured for $\Xi_c^0\to\Xi^0\eta^{(\prime)}$, so the corresponding simulated signal samples are not weighted. 
Background samples of $\Upsilon(4S) \to B^+B^-$ and $B^0\bar{B}^0$ decays at Belle and Belle II, as well as $\Upsilon(5S) \to B_s^{(*)0}\bar{B}_s^{(*)0}$ decays at Belle, are generated using {\sc evtgen} and {\sc pythia}.~The continuum background from $e^+e^- \to q\bar{q}$ processes, where $q$ indicates a $u$, $d$, $c$, or $s$ quark, is generated by the {\sc kkmc}~\cite{kkmc} software package, with {\sc pythia} used for hadronization and {\sc evtgen} for subsequent decays of hadrons.
Final state radiation effects are accounted for using the PHOTOS package~\cite{photos}.
Simulation of the detector response uses the {\sc geant3}~\cite{geant3} and {\sc geant4}~\cite{geant4} software packages for Belle and Belle II, respectively.

\section{Selection criteria}

We reconstruct the decays $\xicz\to\xizpiz$, $\xizeta$, $\xizetap$, and $\xipi$, followed by the decays $\xiz\to\Lambda\piz$, $\xim\to\Lambda\pim$, $\Lambda\to p\pim$, $\etap\to\pip\pim\eta$, and $\piz/\eta\to\gamma\gamma$.
The Belle~II software~\cite{basf2} is used for event reconstruction of both samples, taking advantage of software improvements in Belle~II. 
The Belle data are converted to the Belle II data format~\cite{b2bii}. 
The selection criteria are nearly identical for Belle and Belle~II.
A global decay chain vertex fit is applied for each mode using the TreeFit algorithm~\cite{b2fit}.

For reconstructed charged particles not originating from long-lived $\Xi^-$ and $\Lambda$ baryon decays, the impact parameters, which are the distances of closest approach from the reconstructed trajectory perpendicular to and along the $z$ axis with respect to the nominal interaction point, are required to be less than 0.1~cm and 2~cm, respectively, to suppress misreconstructed tracks and beam background. 
Charged particles are identified using the likelihood ${\mathcal L}_i$ for each particle hypothesis $i$ based on the information provided by the relevant sub-detector systems.
For Belle data, the pion, kaon, and proton particle identification (PID) uses information from the drift chamber, Cherenkov detectors, and the time-of-flight detector~\cite{BellePID1}.
Information from all subdetectors except the pixel detector is used to determine PID likelihoods for Belle II data.

The reconstruction and selection of $\xim$ and $\xiz$ candidates are the same as those in refs.~\cite{omega2018,xic02xi0kk2021,xic02xi0ll2023}, except for the kinematic requirement on candidate $\piz$'s as detailed below.
The $\Lambda$ candidates are reconstructed via the $\Lambda \to p \pi^-$ decay, where the proton is identified by a PID requirement ${\mathcal L}_p/({\mathcal L}_p+{\mathcal L}_{\pi})>0.2$ and ${\mathcal L}_p/({\mathcal L}_p+{\mathcal L}_K)>0.2$ for Belle and ${\mathcal L}_p/({\mathcal L}_p+{\mathcal L}_e+{\mathcal L}_\mu+{\mathcal L}_\pi+{\mathcal L}_K+{\mathcal L}_d)>0.01$ for Belle II, and no PID requirement is applied to the pions.
The selection efficiency of the PID requirement and the probability of misidentifying a hadron, depending on the particle species and kinematic properties, are approximately 90\% (94\%) and 1\% (1\%), respectively, at Belle (Belle II) in this case. 
The invariant mass of the reconstructed $\Lambda$ candidate must be within 3.5~MeV/$c^2$, corresponding to approximately two times the mass resolution, $\sigma$, of the known mass~\cite{pdg}. 
Each $\pi^-$ candidate from the $\Xi^-$ decay is required to have a transverse momentum greater than 50~MeV/$c$ to remove backgrounds from low-momentum pions.
Candidate $\pi^0$'s from $\xiz$ decays are reconstructed from pairs of photons selected from energy deposits in the ECL (clusters). 
To suppress low-momentum and fake photons, each photon candidate is required to have energy greater than: 30~MeV in the ECL barrel region ($-0.63<\cos\theta<0.85$); 
50 (80)~MeV for Belle (Belle~II) in the forward endcap ($0.85<\cos\theta<0.98$);
and 50 (60)~MeV in the backward endcap ($-0.91<\cos\theta<-0.63$), where $\theta$ is the polar angle in the laboratory frame. 
The reconstructed invariant mass of the photon pair is required to be within 11.6~MeV/$c^2$ (approximately $2\sigma$) of the known $\pi^0$ mass. 
The momenta of the $\pi^0$ candidates in the laboratory frame are required to exceed 0.25~GeV/$c$. 
Candidate $\xim$ and $\xiz$ baryons are formed from $\Lambda\pim$ and $\Lambda\piz$ combinations, respectively.
A vertex fit is applied to the entire $\Xi^{-(0)}$ decay chain, including subsequent decay products, with the $p\pi$ and diphoton masses constrained to match the known $\Lambda$ and $\pi^0$ masses~\cite{pdg}.

The reconstructed $\xim$ and $\xiz$ masses are required to be within 6~MeV/$c^2$ and 5~MeV/$c^2$ (approximately 3$\sigma$ and 1.5$\sigma$) of their known masses, respectively.
These selections are optimized by maximizing the figure-of-merit $N_{\rm sig}/\sqrt{N_{\rm sig}+N_{\rm bkg}}$, where $N_{\rm sig}$ and $N_{\rm bkg}$ are the numbers of $\xicz$ signal events and background events in the $\xicz$ signal region.
The $\xicz$ signal regions are the $\xicz$ invariant mass ranges of (2.4, 2.54), 
(2.25, 2.65), (2.3, 2.6), and (2.37, 2.57)
~GeV$/c^2$ for the $\xicz\to\xipi$, $\xicz\to\xizpiz$, $\xicz\to\xiz\eta$, and $\xicz\to\xiz\etap$ decay modes, respectively.
These regions contain more than 95\% of the simulated signals. 
For the normalization mode, $N_{\rm sig}$ and $N_{\rm bkg}$ are obtained via an unbinned extended maximum-likelihood (EML) fit to the $\Xi^-\pi^+$ invariant mass spectrum in data.
For the signal modes, $N_{\rm sig}$ is the number of expected signal events using the branching fraction predictions in ref.~\cite{theory13su3f2023} and $N_{\rm bkg}$ is the number of background events from the simulated samples of size similar to our data. 
The optimized $\Xi^0$ mass requirements do not strongly depend on $h^0$ and assumed branching fractions, hence we use the same mass requirements for all three signal modes.

ECL clusters are used to reconstruct photons to form $\piz$, $\eta$, and $\etap$ candidates from $\xicz$ decays. 
To reduce the background originating from neutral hadrons, we require the energy deposited in a $3\times3$ matrix of crystals centered on the leading-energy crystal to be 80\% or more of the energy deposited in the surrounding $5\times5$ matrix, in which, different with Belle, outer corner crystals are not considered in Belle~II data.
Candidate $\piz$ and $\eta$ mesons are reconstructed by combining pairs of photons, whose energies are required to be greater than 80, 300, and 150~MeV for $\xicz\to\xizpiz$, $\xicz\to\xizeta$, and $\xicz\to\xizetap$, respectively. 
We reconstruct $\etap$ candidates by combining an $\eta$ candidate with a pair of oppositely-charged pions, which must satisfy a PID requirement of ${\mathcal L}_\pi/({\mathcal L}_\pi+{\mathcal L}_{K})>0.2$ with identification efficiencies of 99\% and misidentification probabilities of 1\% for both Belle and Belle II.
Loose mass windows are then used to select $\piz$, $\eta$ or $\etap$ candidates, with ranges of (0.08, 0.18), (0.4, 0.65) and (0.92, 1.0)
~GeV$/c^2$ , respectively. 
A requirement on the kinematic mass-constrained fit quality is applied, $\chi^2<5$ for the signal $h^0$ candidate. 
The momentum in the c.m. frame obtained from the mass-constrained fit for the selected $\hz$ candidate from the $\xicz$ is required to exceed 0.8~GeV/$c$ in order to suppress background with low momentum neutral particles.

The $\Xi^0_c$ candidates are reconstructed either by combining a $\Xi^-$ candidate with a $\pi^+$ candidate, or by combining a $\Xi^0$ candidate with a $\pi^0$, $\eta$, or $\etap$ candidate. 
To identify the $\pi^+$ candidate, we use the selections ${\mathcal L}_\pi/({\mathcal L}_\pi+{\mathcal L}_{K})>0.2$ and ${\mathcal L}_\pi/({\mathcal L}_\pi+{\mathcal L}_{p})>0.2$, with signal efficiencies of 96\% and 94\% for $\xicz$ selection, and misidentification probabilities of 3\% and 2\% for Belle and Belle II, respectively.
To suppress backgrounds, especially those from $B$-meson decays, we require the scaled momentum, $x_p = p^{*}_{\Xi_c^0}c/\sqrt{s/4-M^{2}({\Xi_c^0})c^{4}}$, of the $\xicz$ candidate to be greater than 0.55, where $p^{*}_{\Xi_c^0}$ is the momentum of $\Xi_c^0$ candidate in the c.m.~frame, $s$ is the square of c.m.~energy, and $M({\Xi_c^0})$ is the invariant mass of the $\Xi_c^0$ candidate.
The selection criteria for photon energies, $\hz$ momentum, $\chi^2(\hz)$, and $x_p$ are optimized by maximizing the figure-of-merit as indicated above.
Optimizing the selection criteria with an alternative parameterization $\varepsilon/(5/2+\sqrt{N_{\rm bkg}})$~\cite{PunziFOM}, where $\varepsilon$ is the reconstruction efficiency for $\xicz \to \Xi^0 h^0$, gives consistent results.

The fractions of events that have multiple candidate events in signal simulations are about 2\% (3\%), 6\% (7\%), 6\% (7\%), and 7\% (9\%) for $\xicz\to\xipi$, $\xicz\to\xiz\piz$, $\xicz\to\xiz\eta$, and $\xicz\to\xiz\etap$, respectively in Belle (Belle II) data. 
These values are consistent with the multiple candidate rates observed in the data.
All $\xicz\to\xipi$ candidates are retained after applying these selections.
For $\xicz\to\xizhz$ events with a single $\xiz$ candidate but multiple $\hz$ candidates, the $\hz$ candidate with the minimum mass-constrained fit $\chi^2$ is selected. 
If an event has multiple $\xiz$ candidates, one is selected at random.
This candidate selection procedure yields simulated signal efficiencies for events with multiple candidates of 53\% (46\%), 51\% (42\%, and 54\% (44\%) for $\xicz\to\xiz\piz$, $\xicz\to\xiz\eta$, and $\xicz\to\xiz\etap$, respectively, at Belle (Belle II). 
After this selection, the overall purities in signal regions of the simulated samples increase by 2\% (4\%), 2\% (4\%), and 3\% (5\%) for $\xicz\to\xiz\piz$, $\xicz\to\xiz\eta$, and $\xicz\to\xiz\etap$, respectively in Belle (Belle II) data.

\section{Branching fractions for $\xicz\to\xizpiz$, $\xicz\to\xizeta$, and $\xicz\to\xizetap$}

\begin{figure}[htbp]
	\centering

\includegraphics[width=6.0cm]{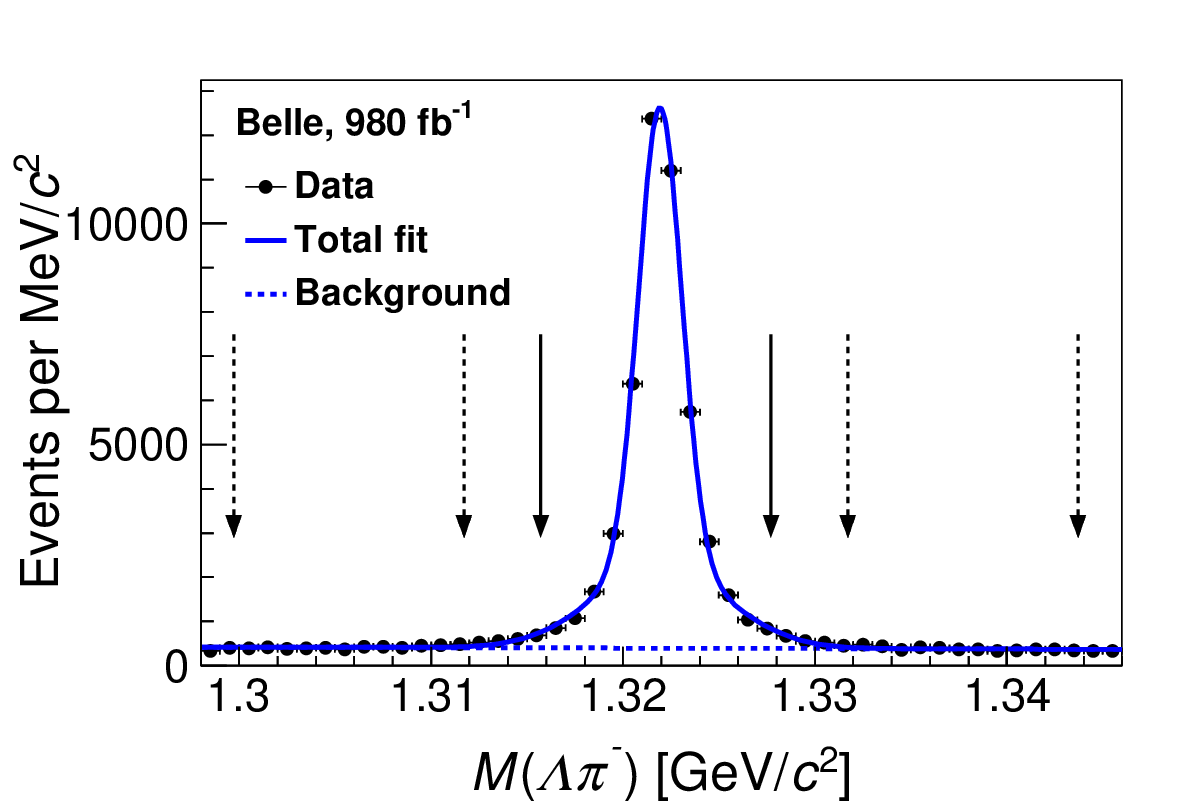}	\put(-40,90){\bf (a)}		
\includegraphics[width=6.0cm]{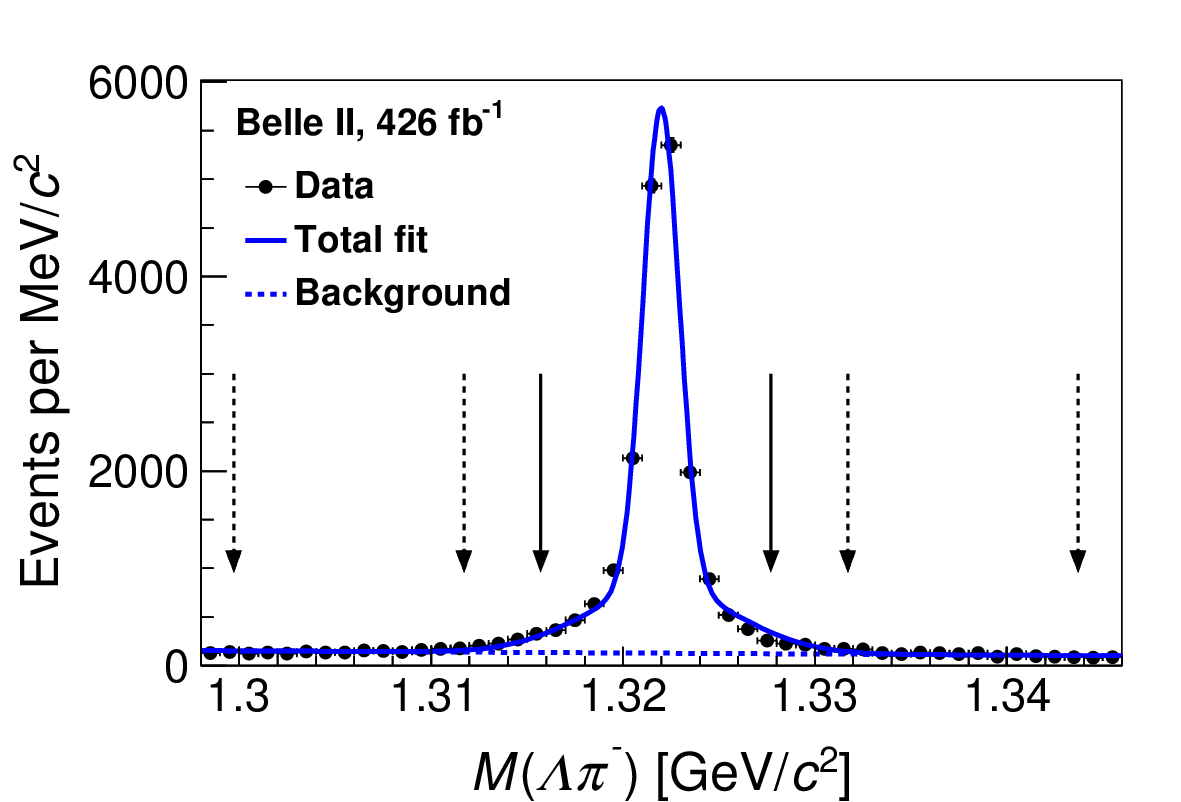}		

\includegraphics[width=6.0cm]{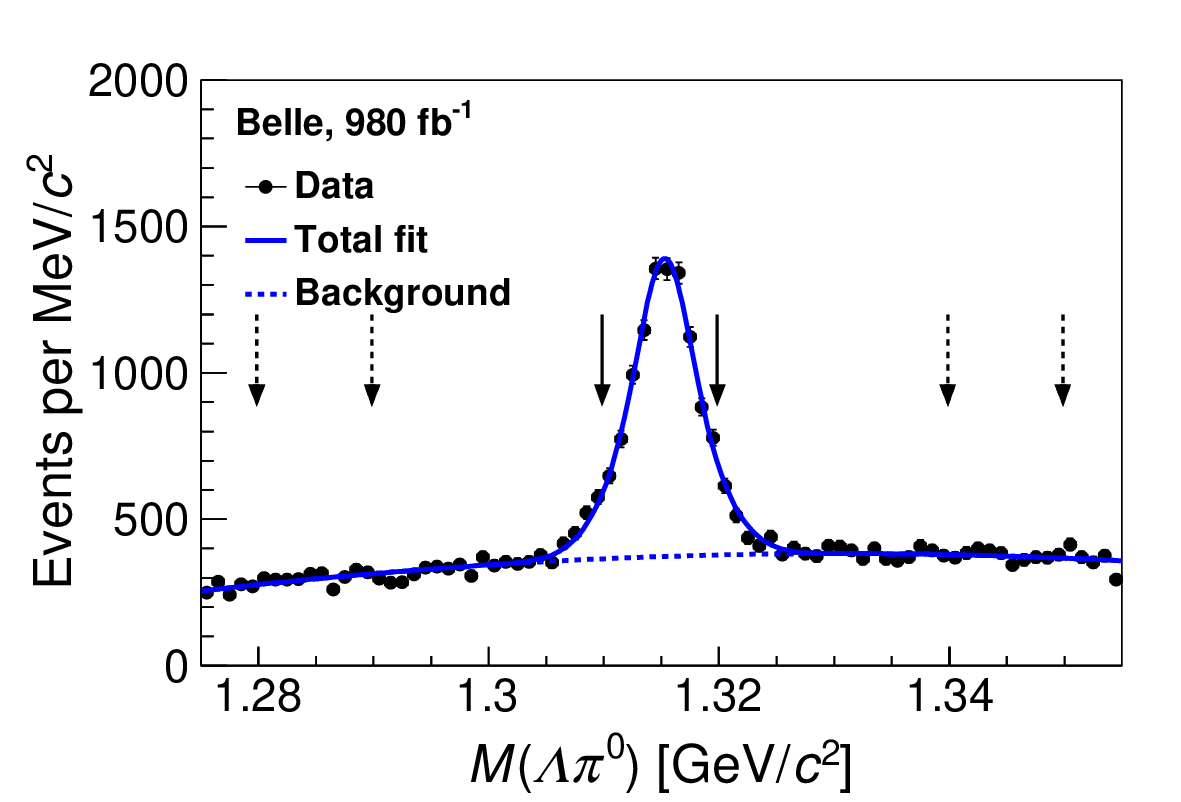}		\put(-40,90){\bf (b)}	
\includegraphics[width=6.0cm]{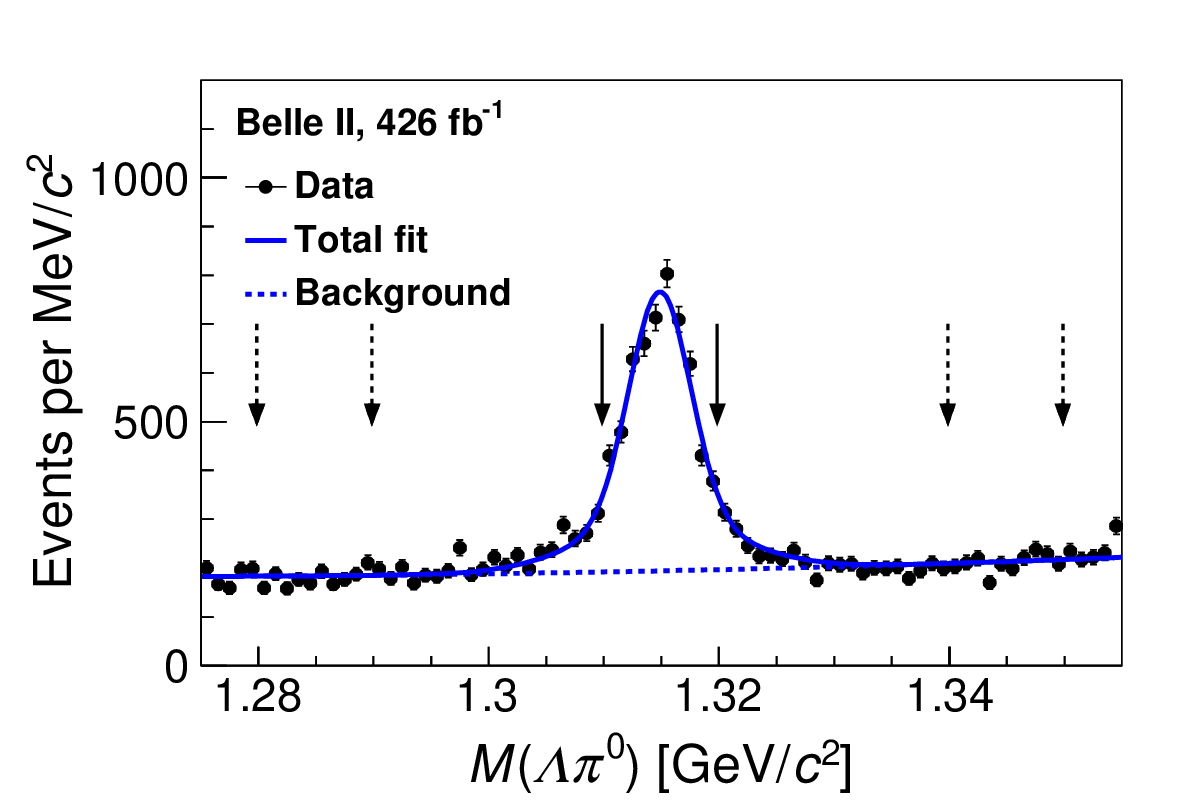}		

\includegraphics[width=6.0cm]{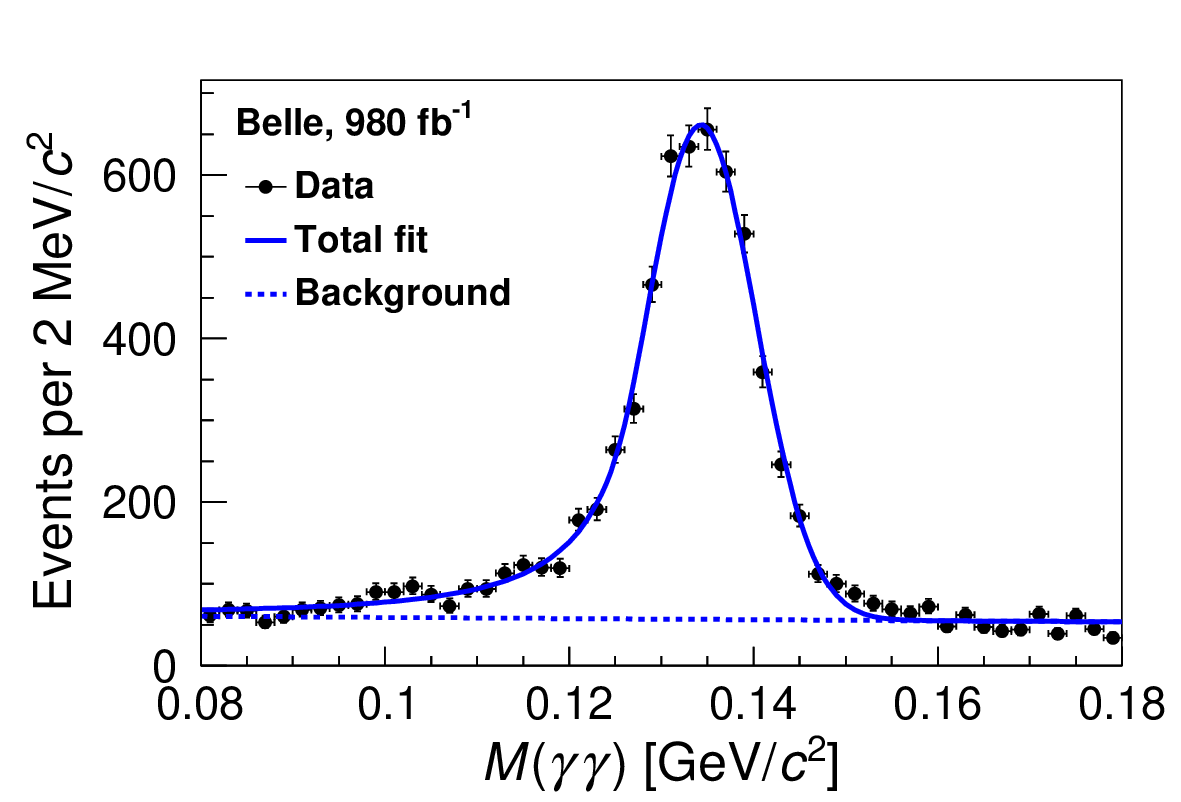}		\put(-40,90){\bf (c)}
\includegraphics[width=6.0cm]{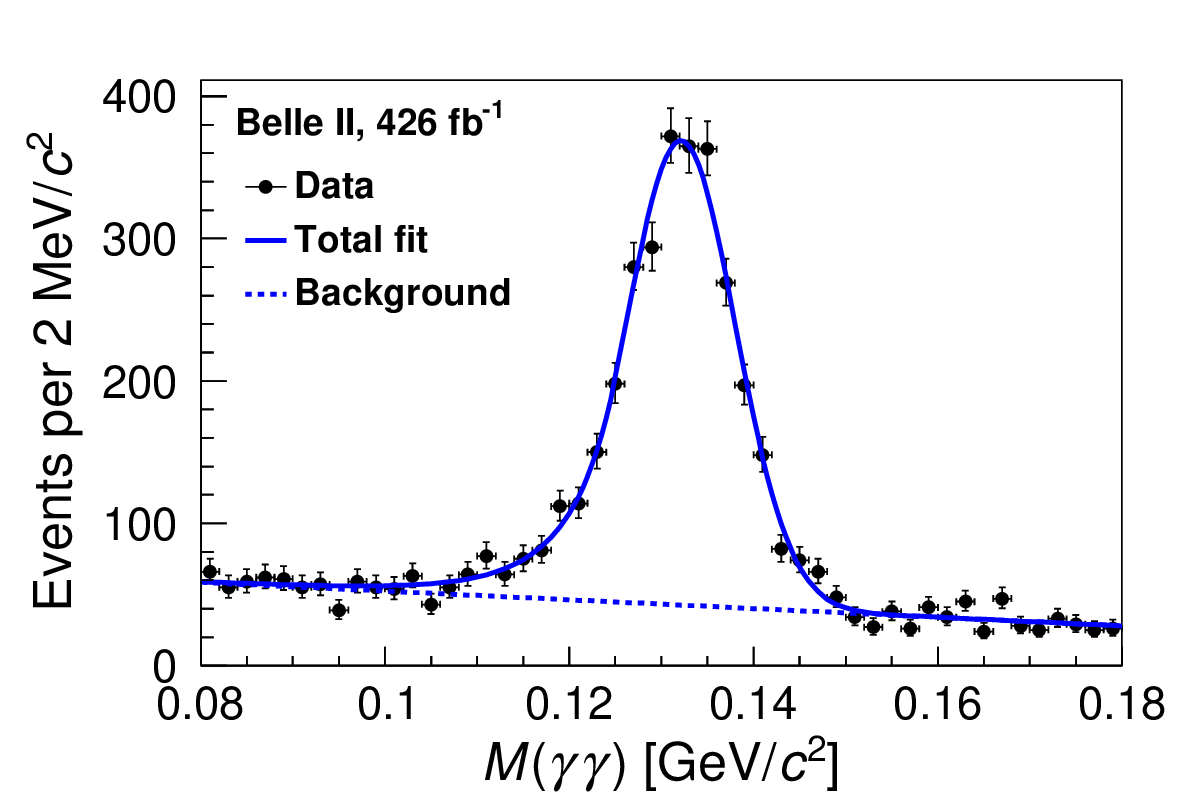}		

\includegraphics[width=6.0cm]{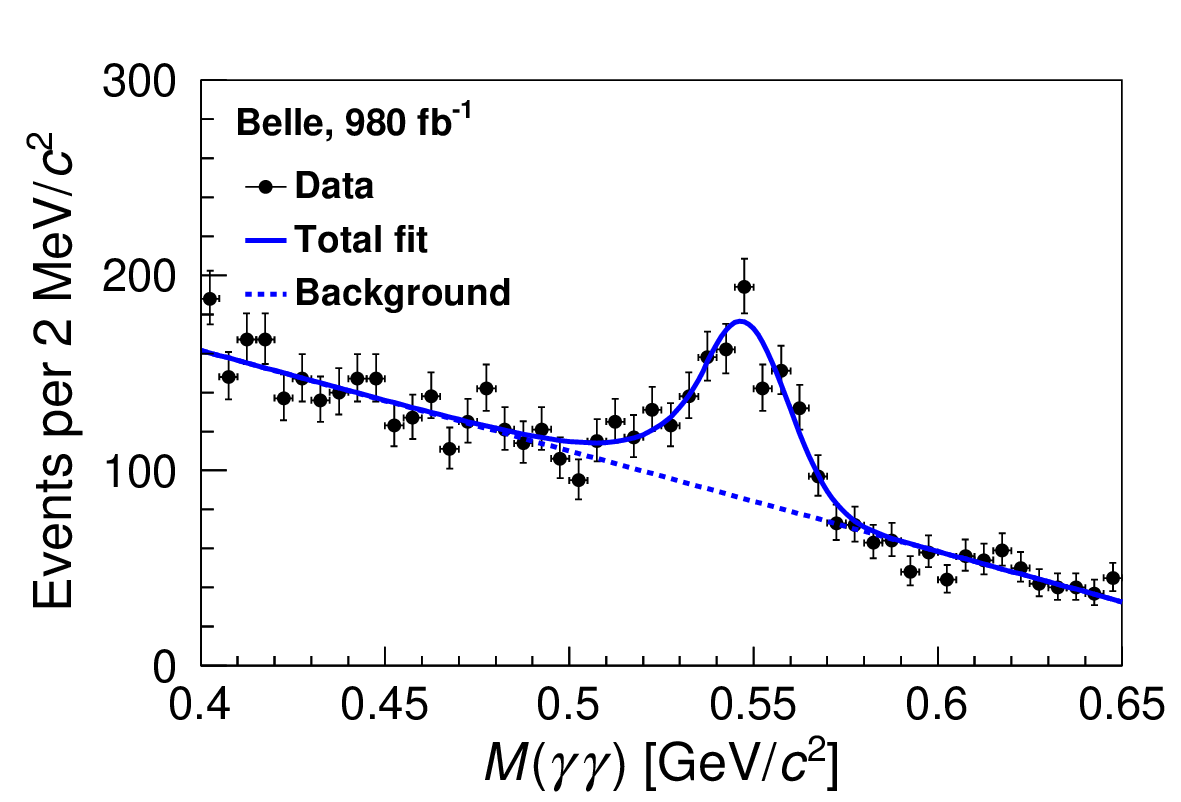}		\put(-40,90){\bf (d)}
\includegraphics[width=6.0cm]{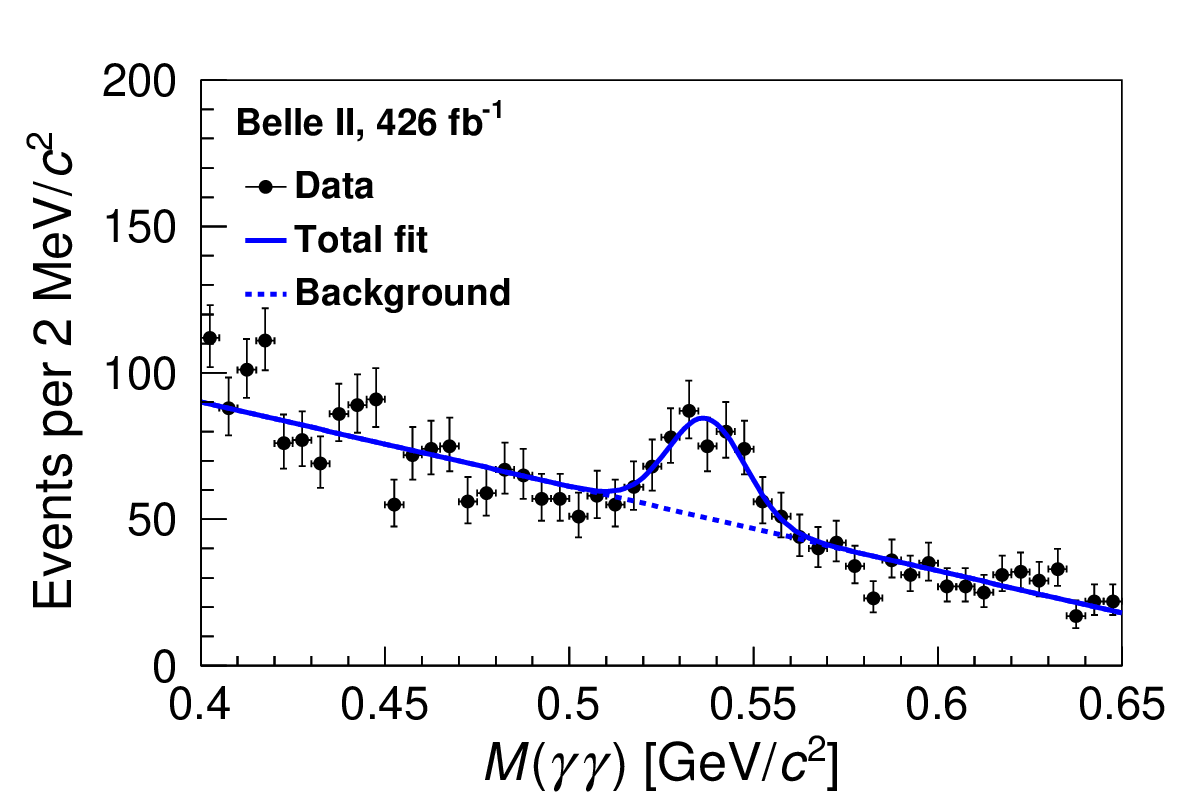}		

\includegraphics[width=6.0cm]{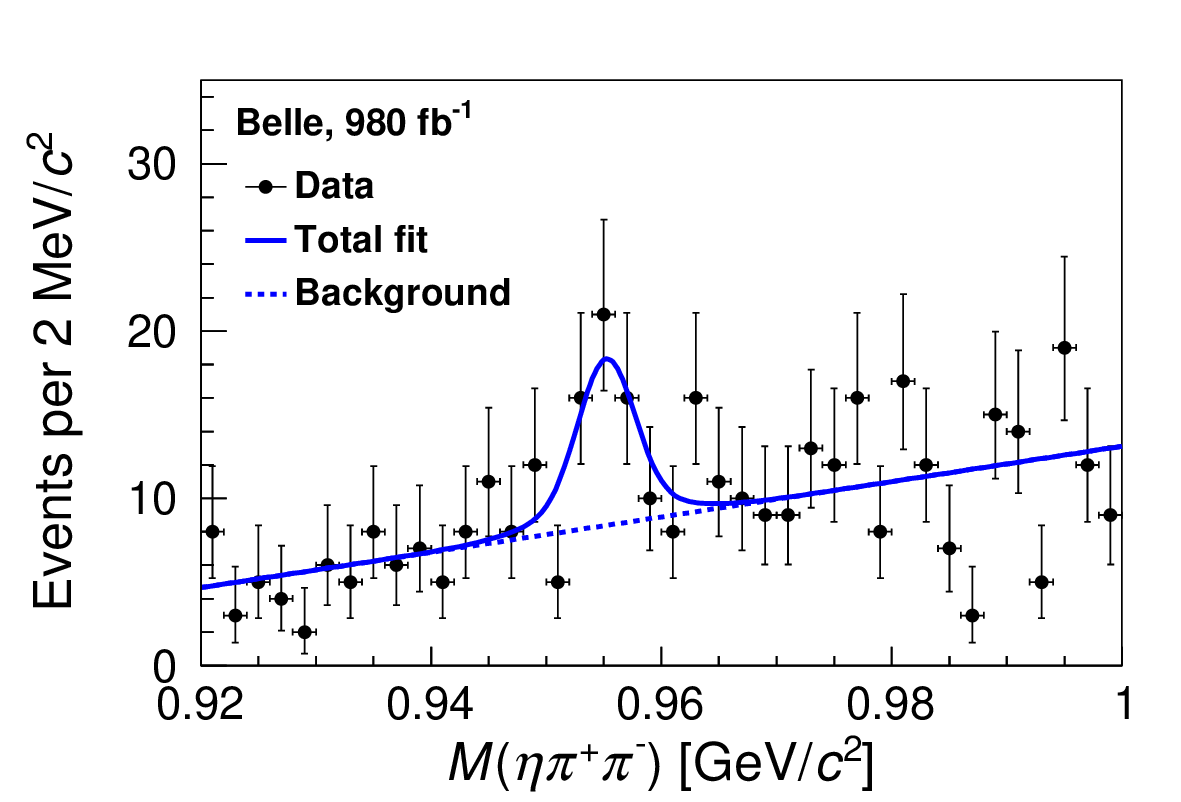}		\put(-40,90){\bf (e)}
\includegraphics[width=6.0cm]{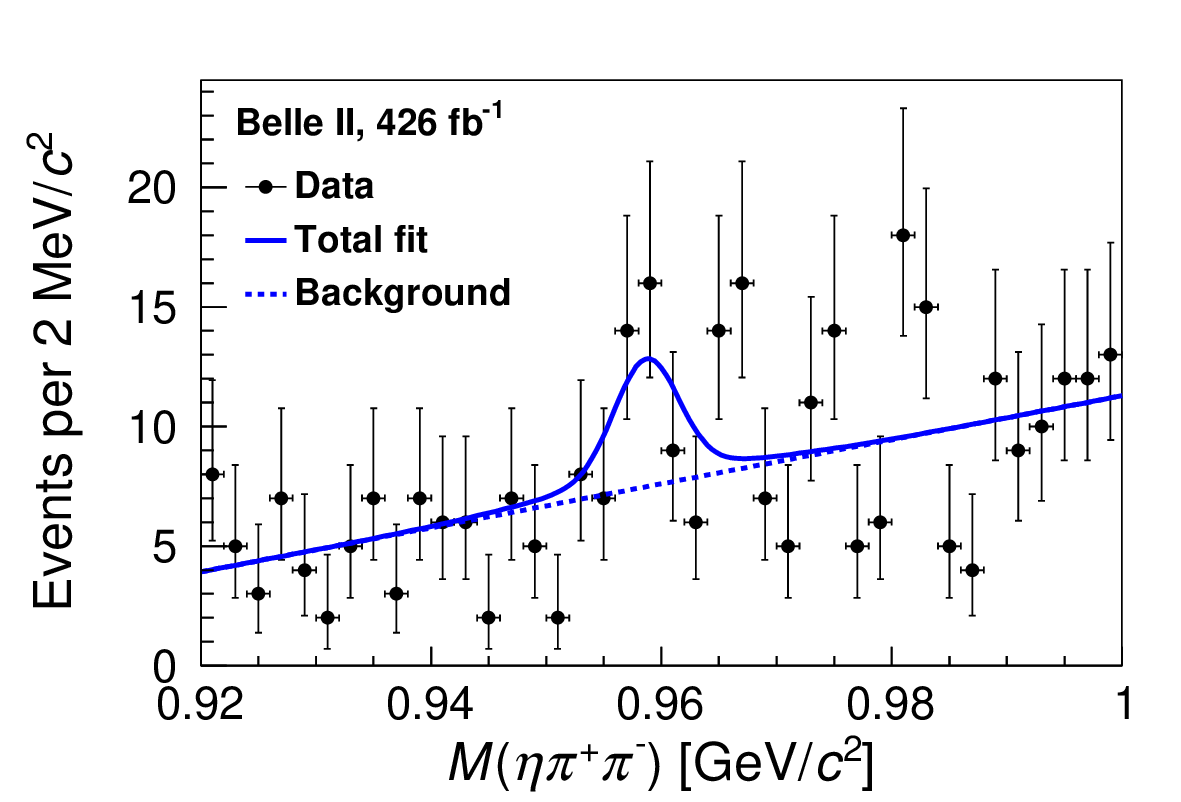}

	\caption{Invariant mass distributions of (a) $\Lambda\pim$ for $\xim$ candidates, (b) $\Lambda\piz$ for $\xiz$ candidates, (c) $\gamma\gamma$ for $\piz$ candidates, (d) $\gamma\gamma$ for $\eta$ candidates, and (e) $\pip\pim\eta$ for $\etap$ candidates in the $\xicz$ signal regions. 
		For each distribution, the left plot shows the result for the Belle sample and the right plot shows that for the Belle II data.
		The markers with error bars represent the data, the solid curves show the total fit, and the dashed curves show the smooth background component of the fit.
		The solid and dashed arrows show the signal and sideband regions for $\Xi^{-(0)}$ candidates, respectively.
	}
	\label{Mintermediate}
\end{figure}

Figure~\ref{Mintermediate} shows the $\Lambda\pim$, $\Lambda\piz$, $\gamma\gamma$, and $\pip\pim\eta$ invariant mass distributions, 
along with the fit results, for $\Xi^0_c$ candidates in the signal region using Belle and Belle II data.
All event selection criteria described in Section~4 are applied, except for the candidate selection procedure and the selection on the corresponding invariant mass region or mass-constrained fit $\chi^2$. 
To illustrate the distributions of the intermediate states, we perform binned EML fits to the invariant mass distributions of the intermediate $\Xi$ and $h^0$ states, where the signal probability density functions (PDFs) are parameterized using a double-Gaussian function with a common mean for the $\xim$, $\xiz$, and $\etap$ candidates, and a Crystal Ball function~\cite{cbfunction} for the $\piz$ and $\eta$ candidates.
The smooth combinatorial backgrounds are described with a straight line for the $\gamma\gamma$, $\eta\pi^+\pi^-$ and $\Lambda \pi^-$ distributions, and a second-order polynomial for the $\Lambda \pi^0$ distributions. 
All the parameters are allowed to float in the fits, except those for the $\etap$ signal shape, which are fixed to the values determined from simulation due to the limited size of the samples.
The solid and dashed arrows indicate the signal and sideband regions, respectively, for $\xim$ and $\xiz$ candidates.
We use candidates in the $\Xi^-$ or $\Xi^0$ signal regions for further analysis of signal extraction and events in the sideband regions as a rough estimation of background.

\begin{figure}[htbp]
	\centering
	\includegraphics[width=7cm]{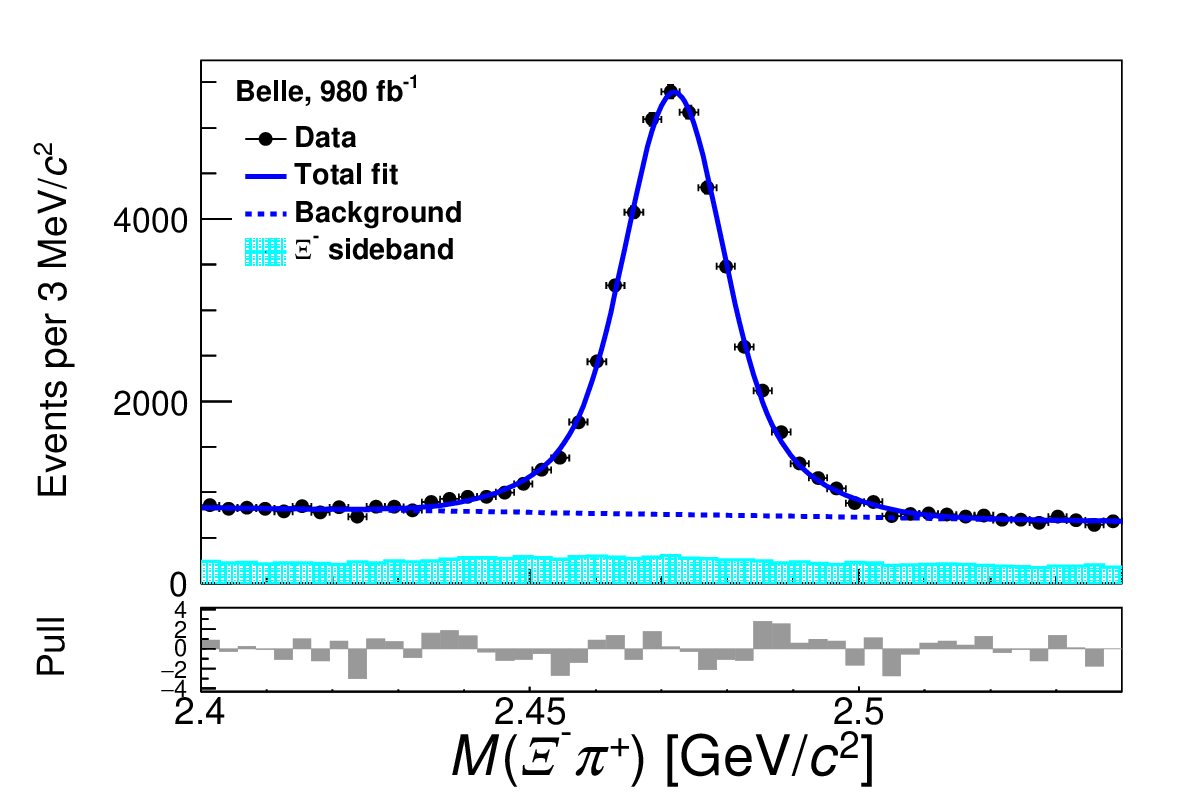}\put(-50,100){\bf (a)}
	\includegraphics[width=7cm]{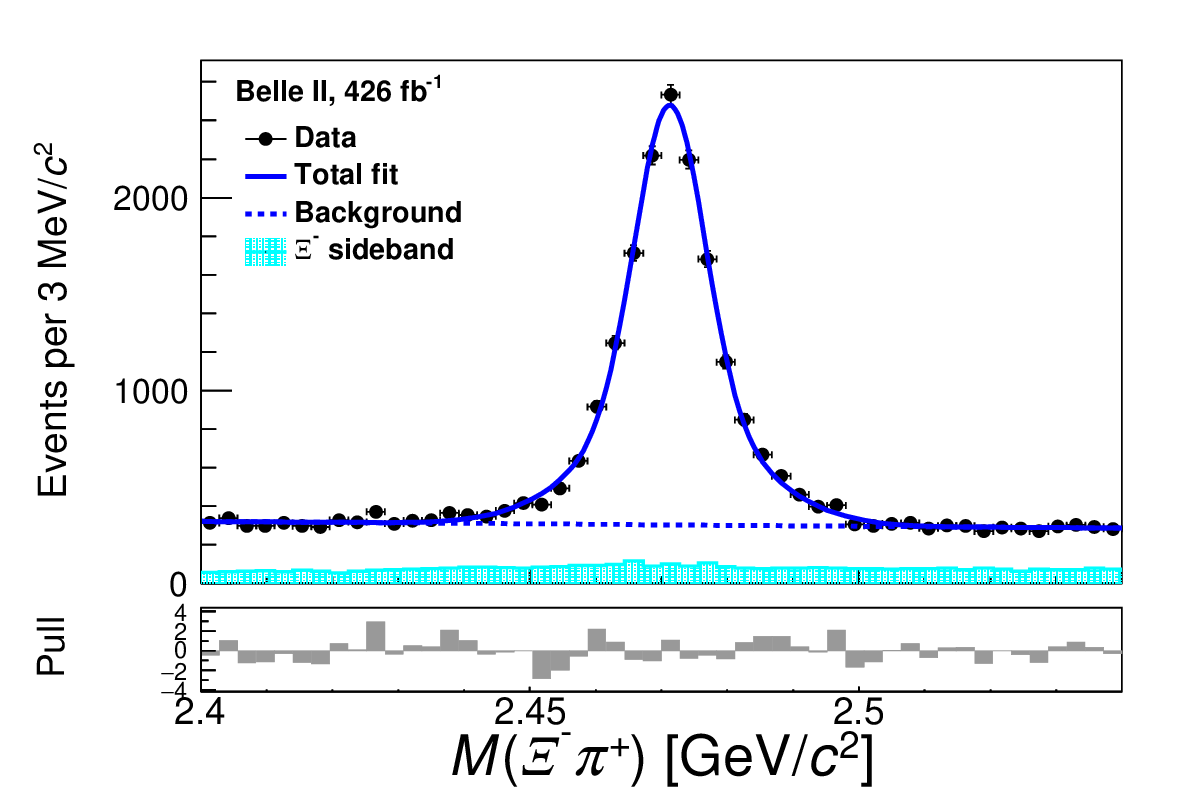}\put(-50,100){\bf (b)}
	\caption{Invariant mass distributions of $\xipi$ from (a) Belle and (b) Belle II data.
The markers with error bars represent the data, the solid blue curves show the fit results, and the dashed blue curves show the background component of the fit.
The cyan histograms are the data from $\xim$ mass sidebands.
}
	\label{Mxic02xipi}
\end{figure}

The $\xipi$ mass distributions for $\xicz$ candidates selected as described in sec.~4 after imposing the $\Xi^-$ signal-mass window requirements are shown in figure~\ref{Mxic02xipi}, together with the results of an unbinned EML fit.
In the fit, the signal shape for $\xicz$ candidates is parameterized by a double-Gaussian function with a common mean and the background shape is described by a straight line. 
All signal and background parameters are floating in the fit. 
The distributions of pulls, $(N_{\rm data}-N_{\rm fit})/\sigma_{\rm data}$, are also displayed in figure~\ref{Mxic02xipi}, where $N_{\rm data}$ is the number of entries in each bin from data, $N_{\rm fit}$ is the fit result in each bin, and $\sigma_{\rm data}$ is the uncertainty on $N_{\rm data}$. 
The fitted signal yields are summarized in table~\ref{Nxic0}.

\begin{figure}[htbp]	
	\centering
	\includegraphics[width=7.0cm]{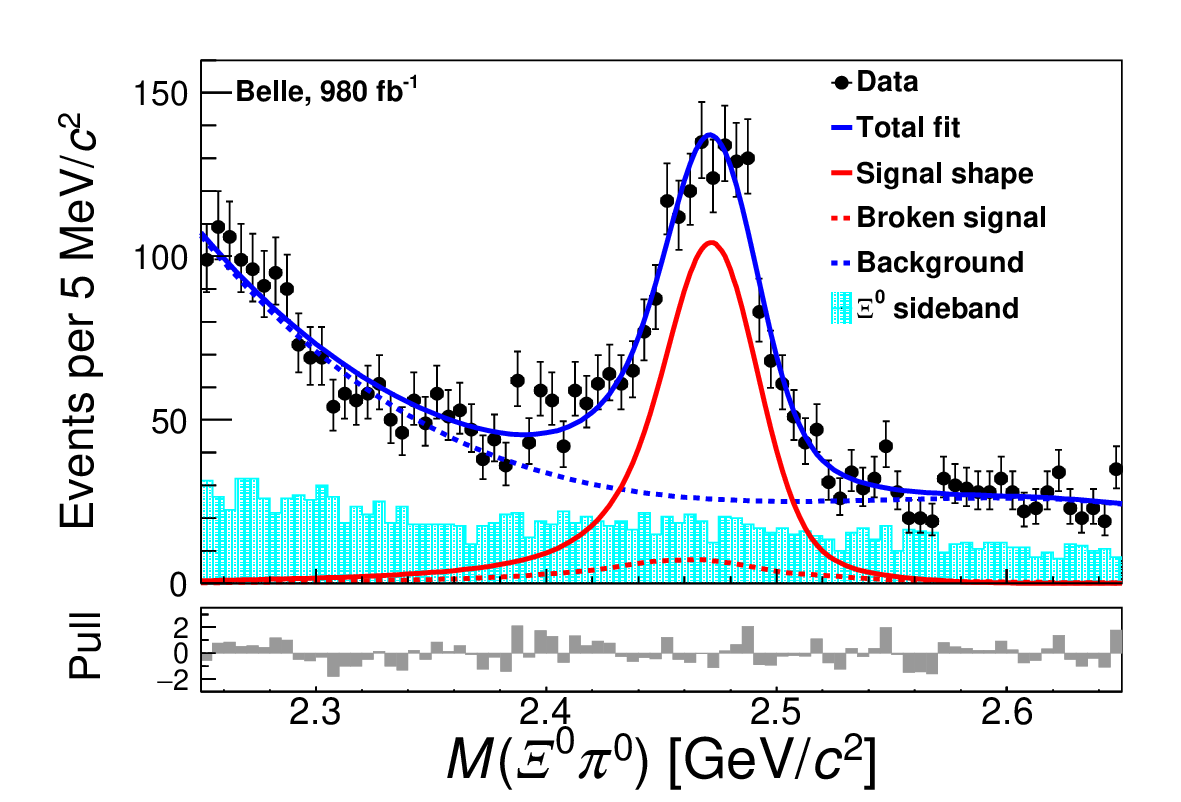}	\put(-150,100){\bf (a)}		
	\includegraphics[width=7.0cm]{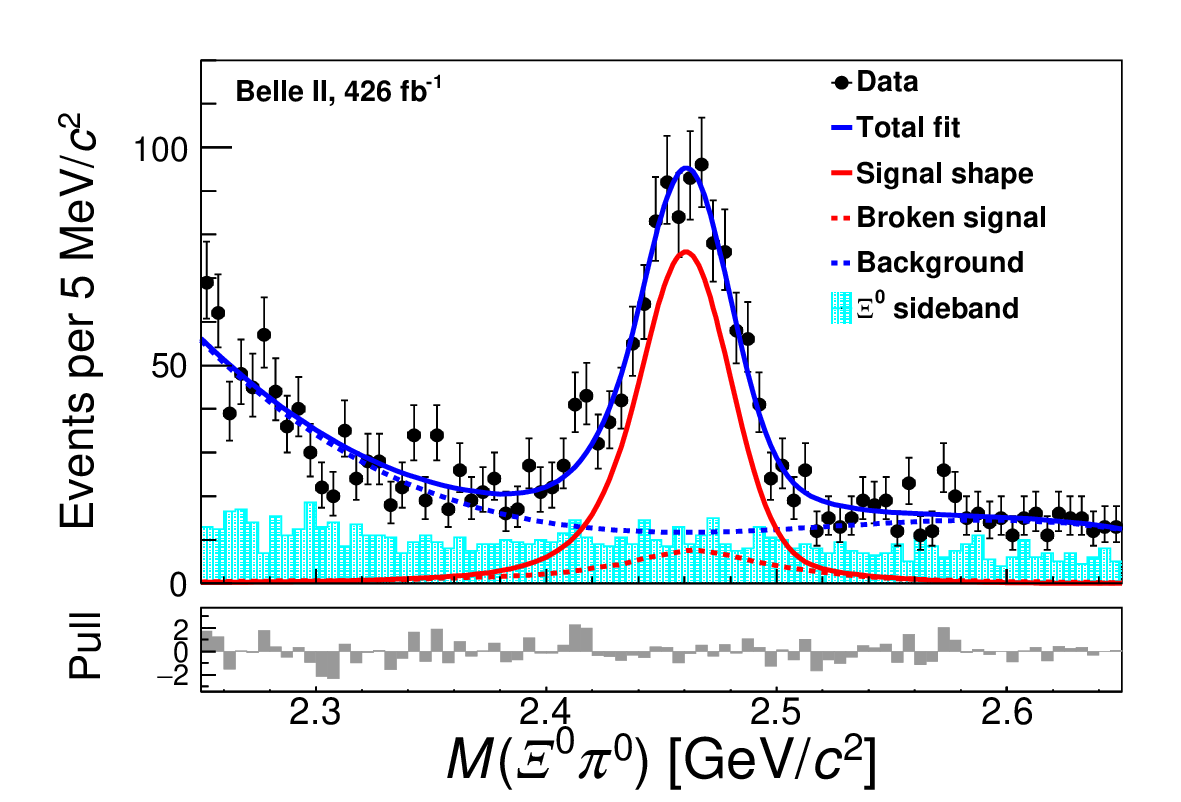}	

	\includegraphics[width=7.0cm]{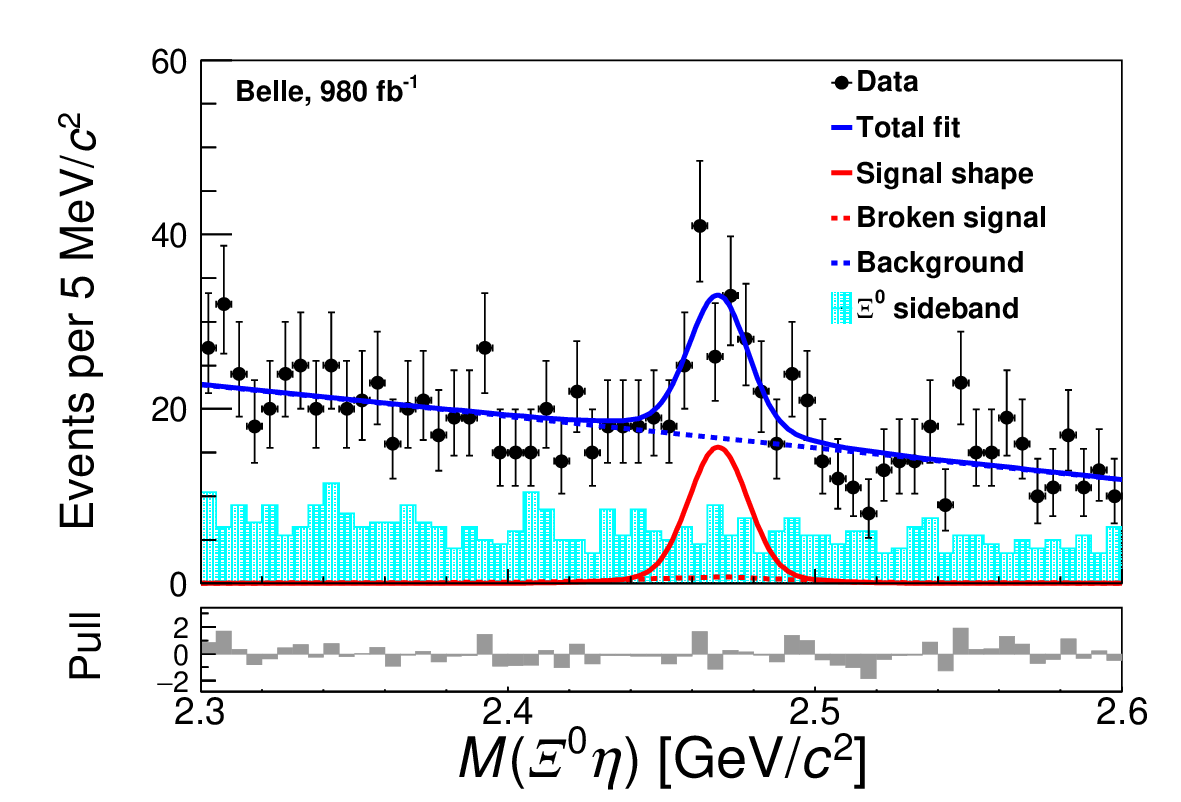}	\put(-150,100){\bf (b)}
	\includegraphics[width=7.0cm]{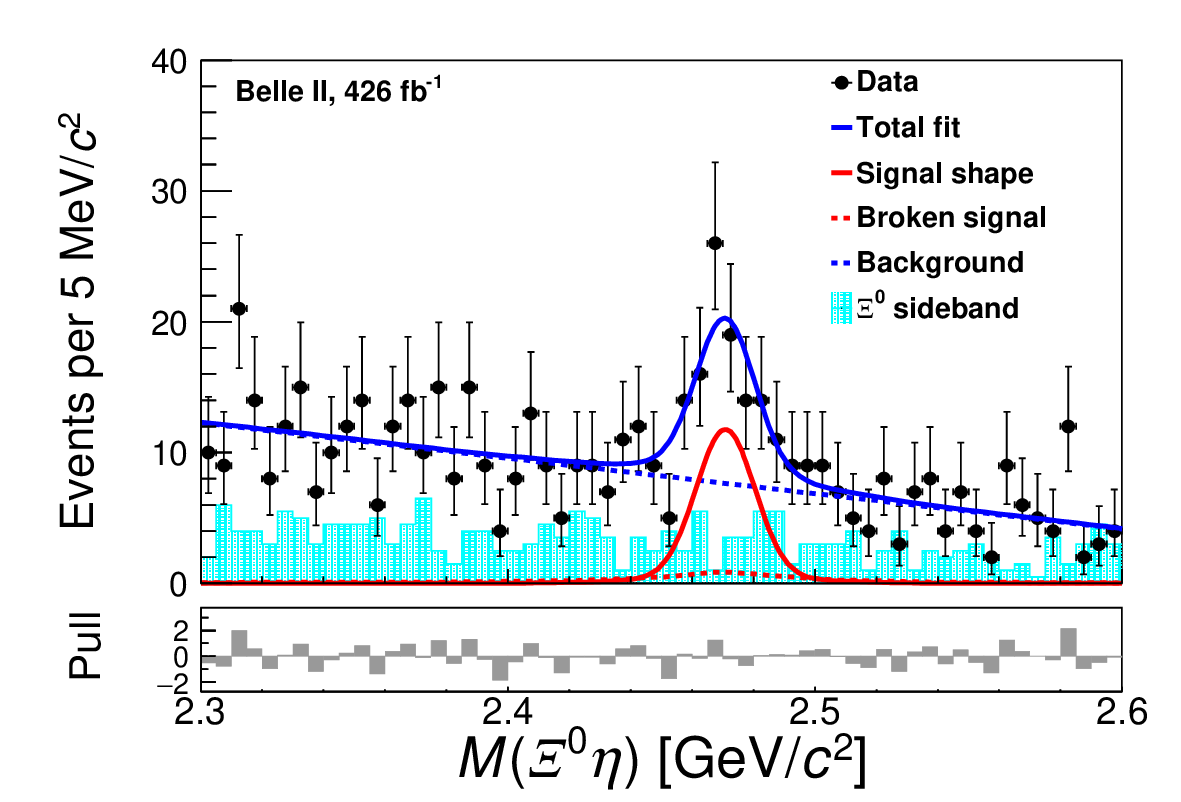}	

	\includegraphics[width=7.0cm]{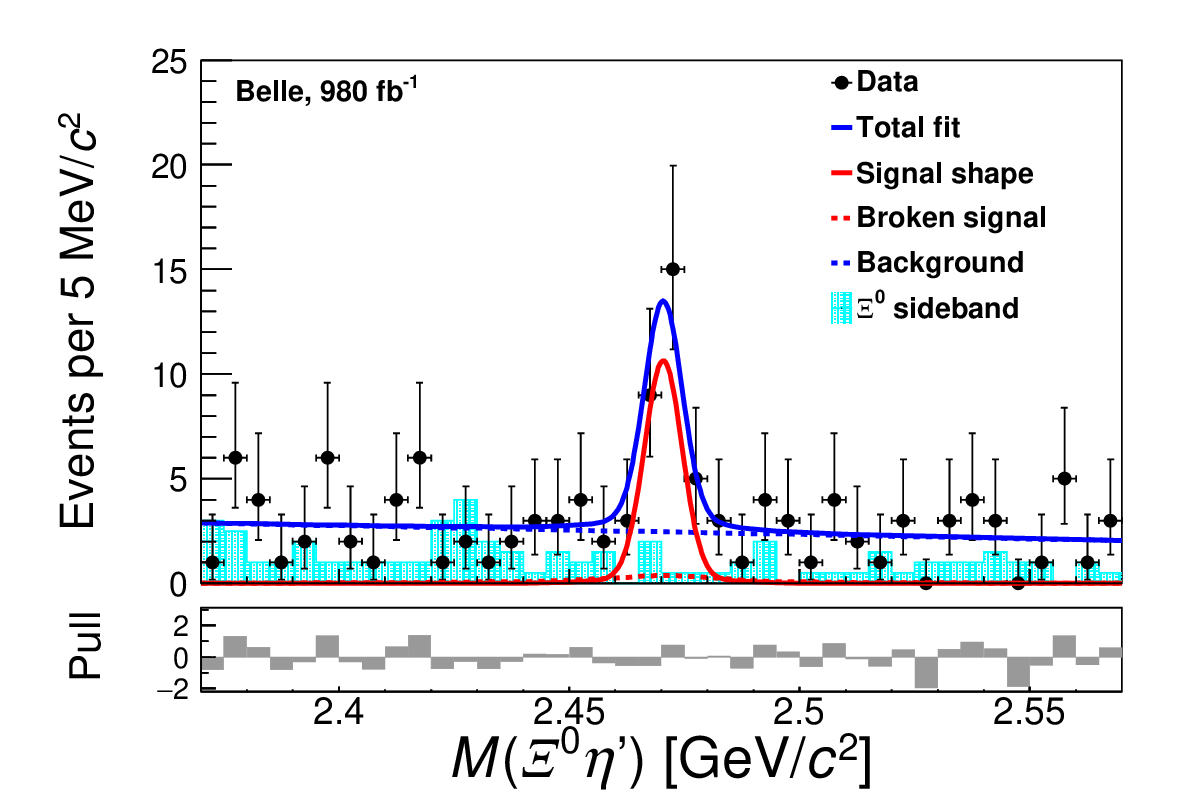}\put(-150,100){\bf (c)}	
	\includegraphics[width=7.0cm]{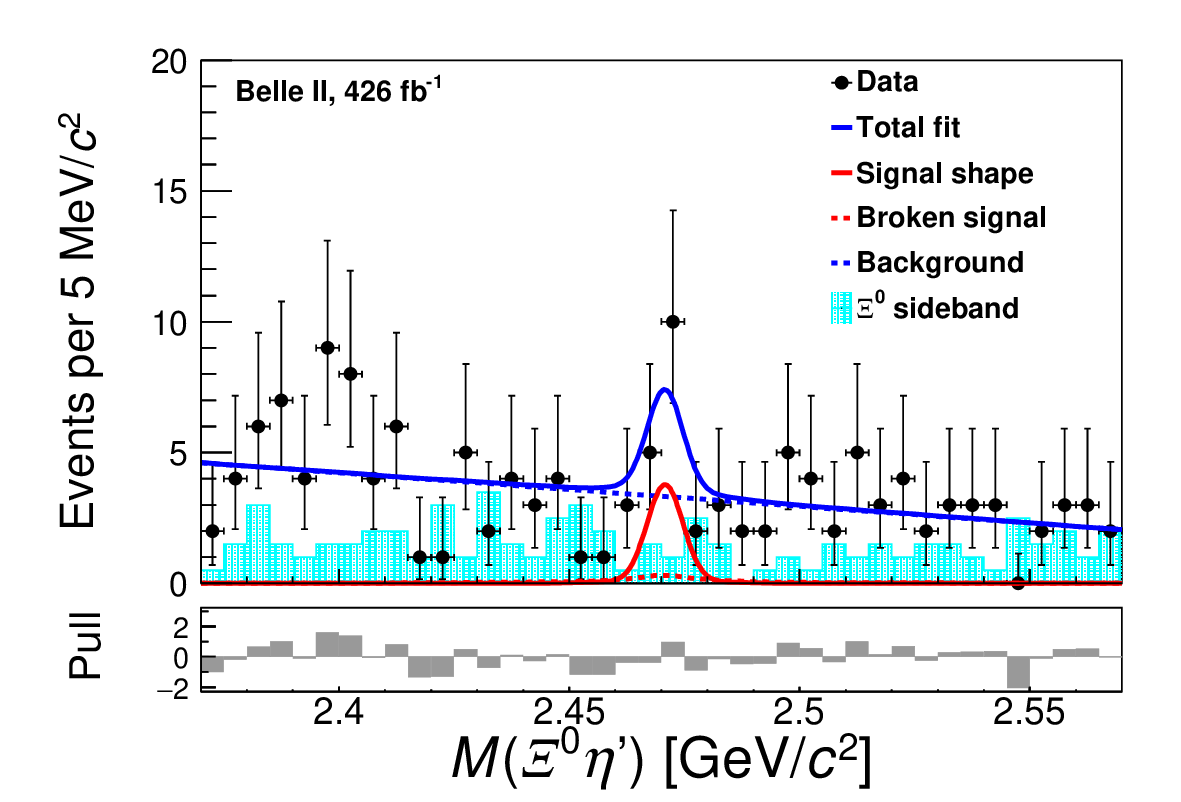}			
	\caption{Invariant mass distributions of $\xicz$ candidates from (a) $\xicz\to\xizpiz$, (b) $\xicz\to\xizeta$, and (c) $\xicz\to\xizetap$ decays reconstructed in (left) Belle and (right) Belle~II data.
The markers with error bars represent the data.
The solid blue curves, solid red curves, dashed red curves, and dashed blue curves show the total fit, signal shape, broken-signal shape, and smooth backgrounds, respectively.
The cyan histograms show the data from the $\xiz$ mass sidebands.
}
	\label{Mxic02xi0h0Data}
\end{figure}

Distributions of $\xizhz$ masses of $\xicz$ candidates reconstructed in data and selected as described in sec.~4 are shown in figure~\ref{Mxic02xi0h0Data} with the results of an unbinned EML fit overlaid. 
The fit PDF includes terms for the signal (${\mathcal F}_{\rm sig}$), broken-signal (${\mathcal F}_{\rm broken}$) and smooth background (${\mathcal F}_{\rm bkg}$) contributions:
\begin{equation}
{\mathcal F}=n_{\rm sig}{\mathcal F}_{\rm sig}+n_{\rm broken}{\mathcal F}_{\rm broken}+n_{\rm bkg}{\mathcal F}_{\rm bkg},
\end{equation}
where $n_{\rm sig}$, $n_{\rm broken}$, and $n_{\rm bkg}$ are the numbers of $\xicz$ signal events, broken-signal events, and smooth background events, respectively.
Here the broken-signal events are those for which at least one of the final state particles, primarily a photon, is not associated with the signal decay. 
The broken-signal events are considered as peaking backgrounds.
The values of $n_{\rm sig}$ and $n_{\rm bkg}$ are allowed to float in the fit, while the ratios of $n_{\rm broken}$ to $n_{\rm sig}$ are fixed to the fractions from signal MC simulation and are 11.6\% (16.0\%), 13.3\% (18.4\%), and 13.3\% (21.0\%) for $\xicz\to\xiz\piz$, $\xicz\to\xiz\eta$, and $\xicz\to\xiz\etap$ decay modes, respectively, at Belle (Belle II). 
Studies based on associating MC simulation generator information to events reconstructed from simulation~\cite{topoana} and $M(\Xi^0h^0)$ distributions from the $\xiz$ and $\hz$ data sidebands show no evidence of peaking backgrounds.
The mass sidebands for $h^0$ are $0.08<M_{\dig}<0.10$ GeV/$c^2$ or $0.16<M_{\dig}<0.18$ GeV/$c^2$ for $\piz$, $0.42<M_{\dig}<0.44$ GeV/$c^2$ or $0.62<M_{\dig}<0.64$ GeV/$c^2$ for $\eta$, and $0.92<M_{\eta\pi^+\pi^-}<0.94$ GeV/$c^2$ or $0.98<M_{\eta\pi^+\pi^-}<1.00$ GeV/$c^2$ for $\eta^{\prime}$, respectively.
The signal PDF in the $\xicz\to\xizpiz$ mode is described by two Crystal Ball functions~\cite{cbfunction} with a common mean, convolved with a Gaussian function to take into account the difference in mass resolution from the simulated events.
For the $\xicz\to\xizeta$ and $\xicz\to\xizetap$ modes, the signal PDF is modeled using double-Gaussian functions with a common mean.
All the signal PDF parameters are fixed to the values obtained from signal simulation, except for the mean value of the signal PDF and the width of the Gaussian resolution function, which are determined in the fit to data.
The width is found to be $11.1 \pm 2.0$ ($8.7 \pm 2.7$)~$\mevcs$ in Belle (Belle~II), where the uncertainty is statistical only.
The ${\mathcal F}_{\rm broken}$ term is a non-parameteric kernel estimation PDF~\cite{rookeyspdf} obtained from simulations.
The ${\mathcal F}_{\rm bkg}$ PDF is parameterized by a third-order polynomial for the $\xicz\to\xizpiz$ mode and by a straight line for the $\xicz\to\xizeta$ and $\xicz\to\xizetap$ modes.
All of the parameters for ${\mathcal F}_{\rm bkg}$ are allowed to vary in the fit. 
Further validation of the fit using simulation confirms that the fit results are unbiased and have Gaussian uncertainties.
The reconstruction efficiencies and fit results are listed in table~\ref{Nxic0}.
The reconstruction efficiencies for $\xicz\to\xizpiz$, $\xicz\to\xizeta$, and $\xicz\to\xizetap$ in Belle II are larger than those in Belle due to improved photon reconstruction stemming from the timing improvements in the ECL readout electronics.
The statistical significances for $\xicz\to\xiz\piz$, $\xicz\to\xiz\eta$, and $\xicz\to\xiz\etap$ are greater than $10\sigma$ ($10\sigma$), $6.2\sigma$ ($6.7\sigma$), and $5.9\sigma$ ($2.4\sigma$) in Belle (Belle II), respectively, calculated using $\sqrt{-2 \ln({\mathcal L}_0/{\mathcal L}_{\rm max})}$, where ${\mathcal L}_0$ and ${\mathcal L}_{\rm max}$ are the maximized likelihoods without and with the signal component, respectively.

	\begin{table}
	\caption{Observed $\xicz$ signal yields and reconstruction efficiencies for various modes, where uncertainties are statistical only.}
	\vspace{0.2cm}
	\centering
	\label{Nxic0}
	\footnotesize
	\begin{tabular}{l c c c c}
		\hline\hline
Mode & Belle yield & $\varepsilon_{\rm Belle}$ (\%) & Belle~II yield & $\varepsilon_{\rm Belle~II}$ (\%) \\
\hline
$\xicz\to\xipi$ & $(363\pm3)\times10^2$ & $13.92\pm0.05$ & $(137\pm2)\times10^2$ & $13.38\pm0.03$ \\
$\xicz\to\xizpiz$ & $1315\pm66$ & $1.09\pm0.01$ & $869\pm46$ & $1.71\pm0.01$ \\
$\xicz\to\xizeta$ & $81\pm15$ & $0.80\pm0.01$ & $60\pm11$ & $1.12\pm0.01$ \\
$\xicz\to\xizetap$ & $23\pm6$ & $0.46\pm0.01$ & $8\pm4$ & $0.81\pm0.01$ \\
		\hline\hline
	\end{tabular}
\end{table}

The ratios of branching fractions to the normalization mode $\xicz\to\xipi$ are calculated via
\begin{equation}\label{bf4}
\begin{split}
& \frac{\BR(\xicz\to\xizpiz)}{\BR(\xicz\to\xipi)}= \frac{N_{\xizpiz}\varepsilon_{\xipi}}{\varepsilon_{\xizpiz}N_{\xipi}}\times\frac{\BR(\xim\to\Lambda\pim)}{\BR(\xiz\to\Lambda\piz)\BR(\piz\to\gamma\gamma)\BR(\piz\to\gamma\gamma)}, \\
& \frac{\BR(\xicz\to\xizeta)}{\BR(\xicz\to\xipi)}= \frac{N_{\xizeta}\varepsilon_{\xipi}}{\varepsilon_{\xizeta}N_{\xipi}}\times\frac{\BR(\xim\to\Lambda\pim)}{\BR(\xiz\to\Lambda\piz)\BR(\piz\to\gamma\gamma)\BR(\eta\to\gamma\gamma)}, \\
& \frac{\BR(\xicz\to\xizetap)}{\BR(\xicz\to\xipi)}= \frac{N_{\xizetap}\varepsilon_{\xipi}}{\varepsilon_{\xizetap}N_{\xipi}}\times\frac{\BR(\xim\to\Lambda\pim)}{\BR(\xiz\to\Lambda\piz)\BR(\piz\to\gamma\gamma)\BR(\etap\to\pip\pim\eta)\BR(\eta\to\gamma\gamma)}. 
\end{split}
\end{equation}
Here, $N_{\xizpiz}$, $N_{\xizeta}$, $N_{\xizetap}$, and $N_{\xipi}$ are the $\xicz$ yields resulting from the fit;
$\varepsilon_{\xizpiz}$, $\varepsilon_{\xizeta}$, $\varepsilon_{\xizetap}$, and $\varepsilon_{\xipi}$ are the corresponding reconstruction efficiencies; and the branching fractions are taken from ref.~\cite{pdg}.
We combine the Belle and Belle II branching fraction ratios and uncertainties using the formulas in ref.~\cite{combine},
\begin{equation}\label{eq:combineBF}
\begin{split}
r=\frac{r_1\sigma_2^2+r_2\sigma_1^2}{\sigma_1^2+\sigma_2^2+(r_1-r_2)^2\epsilon_r^2}, \\
\sigma=\sqrt{\frac{\sigma_1^2\sigma_2^2+(r_1^2\sigma_2^2+r_2^2\sigma_1^2)\epsilon_r^2}{\sigma_1^2+\sigma_2^2+(r_1-r_2)^2\epsilon_r^2}},
\end{split}
\end{equation}
where $r_i$, $\sigma_i$ and $\epsilon_r$ are the branching fraction ratio, uncorrelated uncertainty, and relative correlated systematic uncertainty from each data sample, respectively.
The branching fraction ratios are summarized in table~\ref{BRxic0}, where the first and second uncertainties are statistical and systematic, respectively.
The systematic uncertainties are discussed in detail below.

	\begin{table}
	\caption{Branching fraction ratios of $\xicz\to\xizhz$ decays, where the first and second uncertainties are statistical and systematic, respectively.
}
	\vspace{0.2cm}
	\centering
	\label{BRxic0}
	\footnotesize  
	\begin{tabular}{l c c c}
		\hline\hline
		Mode & Belle & Belle II & Combined \\
		\hline
$\BR(\xicz\to\xizpiz)/\BR(\xicz\to\xipi)$  & $0.47\pm 0.02\pm 0.03$ & $0.51\pm 0.03\pm 0.05$ & $0.48\pm 0.02\pm0.03$ \\
$\BR(\xicz\to\xizeta)/\BR(\xicz\to\xipi)$  & $0.10\pm 0.02\pm 0.01$ & $0.14\pm 0.02\pm 0.02$ & $0.11\pm 0.01\pm0.01$ \\
$\BR(\xicz\to\xizetap)/\BR(\xicz\to\xipi)$ & $0.12\pm 0.03\pm 0.01$ & $0.06\pm 0.03\pm 0.01$ & $0.08\pm 0.02\pm0.01$ \\
		\hline\hline
	\end{tabular}
\end{table}

\section{Asymmetry parameter of $\xicz\to\xizpiz$}

\begin{figure}[htbp]	
	\centering	
	\includegraphics[width=7cm]{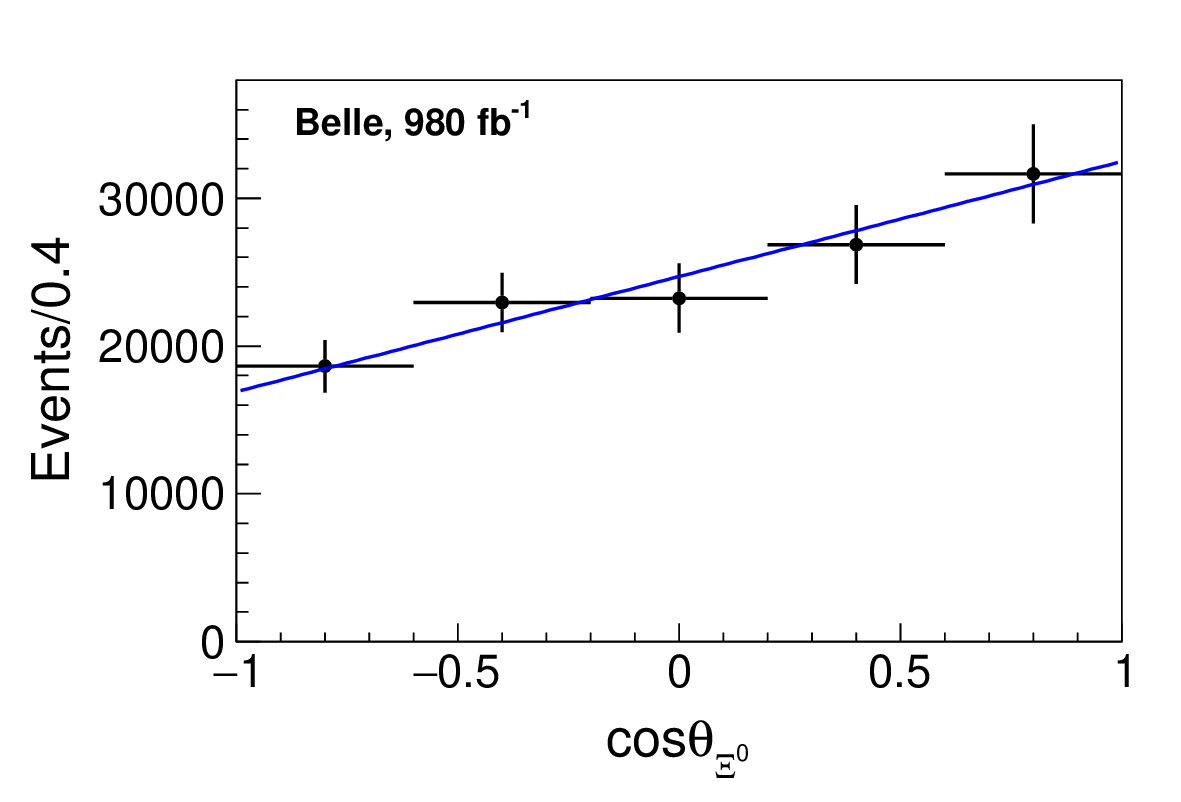} 		\put(-150,100){\bf (a)}	
	\includegraphics[width=7cm]{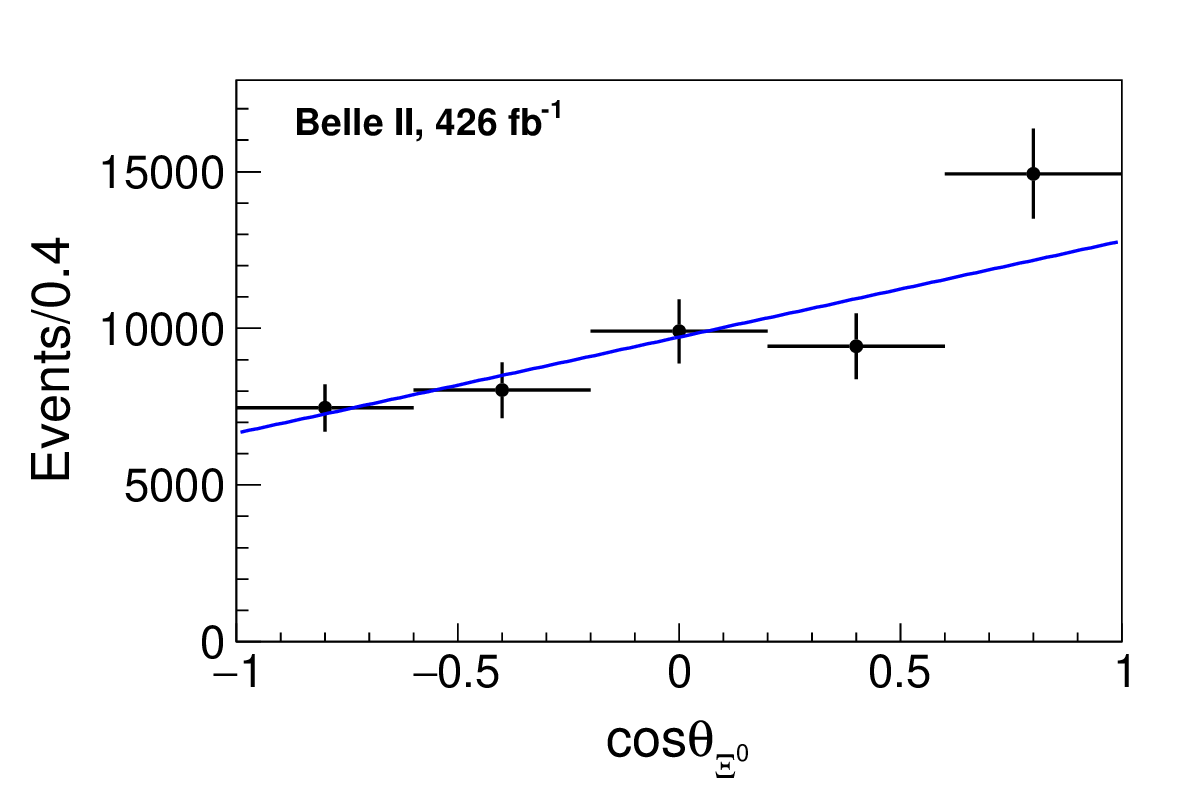} 		\put(-150,100){\bf (b)}
	\caption{Efficiency-corrected $\xicz$ signal yields in bins of $\cos\theta_{\xiz}$ from the (a) Belle and (b) Belle II datasets.
		The lines show linear regression results.
	}
	\label{AlphaXi0pi0}
\end{figure}

Given the small sample sizes for the other modes, the asymmetry parameter is measured only for $\xicz\to\xizpiz$.
We divide the $\cos\theta_{\xiz}$ distribution into five equal sized non-overlapping contiguous intervals (bins).
The $\xicz$ signal yield in each bin is obtained by fitting to the $M(\xiz\piz)$ distribution where the signal shape in each bin is fixed to the corresponding MC simulation and convolved with the Gaussian resolution function, whose width is fixed to the result of a fit to the full sample, due to the limited sample size.
The fits to $M(\xiz\piz)$ spectra in $\cos\theta_{\xiz}$ bins are shown in appendix~\ref{alphaFits}.
Table~\ref{Alpha} lists the signal yields and reconstruction efficiencies in each $\cos\theta_{\xiz}$ bin.
The final efficiency-corrected $\xicz$ signal yields in bins of $\cos\theta_{\xiz}$ for $\xicz\to\xizpiz$ are shown in figure~\ref{AlphaXi0pi0}, together with the simultaneous fit result using eq.~(\ref{eq:1}) with a common value of the product $\alpha(\xicz\to\xizhz)\alpha(\xiz\to\Lambda\piz)$ for the Belle and Belle II data samples.
Using simplified simulated experiments generated with different $\alpha$ values, we test the $\alpha$ extraction procedure and find that it is unbiased. 
The product of asymmetry parameters is found to be $\alpha(\xicz\to\xizpiz)\alpha(\xiz\to\Lambda\piz)=0.32\pm0.05({\rm stat})$.
Taking $\alpha(\xiz\to\Lambda\piz)$ = $-0.349\pm0.009$~\cite{pdg}, we find $\alpha(\xicz\to\xizpiz)=-0.90\pm0.15({\rm stat})\pm0.23({\rm syst})$, where the first uncertainty is statistical and the second is systematic. 
The values of $\alpha(\xicz\to\xiz\piz)$ extracted via individual fits to the Belle and Belle II data samples are $-0.84\pm0.21$ and $-0.98\pm0.22$, where the uncertainties are statistical only, in good agreement with the result from the simultaneous fit.

	\begin{table}
	\caption{Values of the signal yield divided by reconstruction efficiency~(\%) in $\cos\theta_{\xiz}$ bins for $\xicz\to\xizpiz$ in the Belle and Belle~II datasets. }
	\centering
	\vspace{0.2cm}
	\label{Alpha}
	\Large 
	\begin{tabular}{l c c c c c }
		\hline\hline
{\footnotesize $\cos\theta_{\xiz}$} & {\footnotesize $(-1.0, -0.6)$} & {\footnotesize $(-0.6, -0.2)$} & {\footnotesize $(-0.2, 0.2)$} & {\footnotesize $(0.2, 0.6)$} & {\footnotesize $(0.6, 1.0)$}  \\
		\hline
{\footnotesize  Belle} & { $\frac{{260\pm 25}}{1.40}$} & $\frac{{296\pm 26}}{1.29}$ & $\frac{{266\pm 27}}{1.14}$ & $\frac{{265\pm 27}}{0.99}$ & $\frac{{224\pm 24}}{0.71}$ \\
{\footnotesize Belle~II} & { $\frac{{176\pm 18}}{2.37}$} & $\frac{{167\pm 18}}{2.08}$ & $\frac{{194\pm 20}}{1.96}$ & $\frac{{151\pm 17}}{1.60}$ & $\frac{{176\pm 17}}{1.18}$ \\
		\hline\hline
	\end{tabular}
\end{table}

\section{Systematic uncertainties}

\subsection{Branching fraction ratios}
The sources of systematic uncertainties for the branching fraction ratio measurements include those related to the efficiency, the intermediate branching fractions, and the fit procedure.
Table~\ref{BRsyst} summarizes the systematic uncertainties, where the total uncertainty is determined from a quadratic sum of the uncertainties from each source.

	\begin{table}
		\centering
	\caption{Fractional systematic uncertainties (\%) on the relative branching-fraction results. 
		The uncertainties in the last two rows, due to intermediate branching fractions and background shape, are common to Belle and Belle~II; the other uncertainties are independent.
		Since the $\Lambda\to p\pim$ decay is reconstructed in each decay mode, the $\BR(\Lambda\to p\pim)$ uncertainty and the uncertainty due to the $\Lambda\to p\pim$ reconstruction efficiency cancel in the ratio to the reference mode $\xicz\to\xipi$.
}
	\vspace{0.2cm}
	\label{BRsyst}
	\footnotesize
	\begin{tabular}{lcccccc}
		\hline\hline
		\multirow{2}{*}{Source}  & \multicolumn{2}{c}{~~~$\frac{\BR(\xicz\to\xizpiz)}{\BR(\xicz\to\xipi)}$~~~} & \multicolumn{2}{c}{~~~$\frac{\BR(\xicz\to\xizeta)}{\BR(\xicz\to\xipi)}$~~~}  & \multicolumn{2}{c}{~~~$\frac{\BR(\xicz\to\xizetap)}{\BR(\xicz\to\xipi)}$~~~}   \\
		& Belle & Belle~II& Belle & Belle~II& Belle & Belle~II\\
			\hline
		Tracking 						& 0.7 & 0.8 & 0.7 & 0.7 & 1.0 & 1.5 \\
		$\pi^\pm$ PID 					& 0.4 & 0.2 & 0.4 & 0.2 & 1.4 & 0.2 \\
		$\piz$ reconstruction 			& 4.4 & 8.8 & 2.3 & 4.3 & 2.3 & 4.2 \\
		Photon reconstruction			& -	  & -   & 4.0 & 2.0 & 4.0 & 1.9 \\
		Simulation sample size		 	& 0.8 & 0.7 & 0.9 & 0.9 & 1.2 & 1.0 \\
		$\alpha$ uncertainty  		  	& 1.1 & 1.2 & 3.0 & 3.4 & 1.0 & 3.5\\	
		$\xiz$ signal mass window     	& 0.5 & 2.0 & 0.5 & 2.0 & 0.5 & 2.0 \\
		Normalization mode sample size  & 1.0 & 1.3 & 1.0 & 1.3 & 1.0 & 1.3 \\
        Broken-signal ratio ($n_{\rm broken}/n_{\rm sig}$) 	& 2.1 & 1.5 & 3.5 & 3.6 & 3.6 & 5.7\\
        Broken-signal PDF			 	& 0.2 & 0.1 & 7.3 & 7.5 & 2.0 & 1.1 \\
        Mass resolution 	 			& -   & -   & 7.2 & 7.0 & 2.4 & 1.4 \\
        Intermediate states $\BR$       & -   & -   & 0.5 & 0.5 & 1.3 & 1.3 \\
        Background shape				& 4.9 & 4.9 & 9.2 & 9.2 & 6.8 & 6.8 \\
		\hline
		Total  				          	& 7.2 & 10.6&15.3 &15.6 & 9.9 & 11.2 \\
		\hline\hline
\end{tabular}
\end{table}

The systematic uncertainty due to the efficiency includes effects due to the detection efficiency, simulation sample size, $\alpha$ uncertainty, and the mass window for the $\xiz$ signal.
The detection efficiencies determined in simulations are corrected by multiplicative data-to-simulation ratios determined from control data samples.
The correction factors and uncertainties include those from track-finding efficiency, obtained from the control samples of $D^{*+}\to D^0(\to K_S^0 \pi^+ \pi^- )\pi^+$ at Belle and $\overline B^0\to D^{*+}(\to D^{0}\pi^+)\pi^-$ and $e^+e^-\to\tau^+\tau^-$ at Belle II; 
charged pion identification, obtained from the $D^{*+} \to D^0(\to K^-\pi^+)\pi^+$ control sample at Belle and Belle II; 
$\piz$ reconstruction, obtained from the $\tau\to\pi^-\pi^0\nu_\tau$ control sample at Belle and the $D^0\to K^-\pi^+\piz$ control sample at Belle II; 
and photon reconstruction, obtained from control samples of radiative Bhabhas at Belle and radiative muon-pairs at Belle II.
We use the control samples to obtain the ratio of data-to-simulation efficiencies as a function of momentum and polar angle and re-weight these ratios according to the kinematic distributions of signal modes from simulation~\cite{BellePID1,PIDBelle2}. 
Correction factors are determined based on the weighted ratios and the uncertainties on the correction factors are taken as systematic uncertainties.
The uncertainty due to photon reconstruction is included in the uncertainty due to $\piz$ reconstruction.
The relative systematic uncertainty due to the size of the simulated sample is calculated using a binomial uncertainty estimate.
For the $\xizpiz$ channel, we use the largest change in efficiency due to variations of the measured value of $\alpha$ by one standard deviation as a systematic uncertainty; for the other channels we use the largest difference in efficiencies observed when assuming the extreme values $\alpha=+1$ or $-1$. 
The uncertainty due to the $\xiz$ signal region choice is calculated from the difference between the selected signal fractions in simulation and data. 
Since the $\chi^2(\hz)$ distributions from sideband-substracted data and simulations are consistent, the efficiency differences on the $\chi^2<5$ requirement between data and simulations are less than 1\%, and thus the uncertainty due to the $\chi^2$ criterion is neglected here.

The systematic uncertainties due to the intermediate branching fractions are taken to be the uncertainties on the world-average values and treated as correlated uncertainties, which are common to Belle and Belle~II.
Only the uncertainties for $\BR(\eta\to\gamma\gamma)$ (0.5\%) and $\BR(\etap\to\pip\pim\eta)$$\BR(\eta\to\gamma\gamma)$ (1.3\%) contribute for $\xicz\to\xizeta$ and $\xicz\to\xizetap$, respectively.
The uncertainties for other intermediate branching fractions are smaller than 0.1\% and are neglected.
The 22.4\% uncertainty on $\BR(\xicz\to\xipi)$ is treated as an independent systematic uncertainty in the measurement of the absolute branching fractions.

The uncertainties due to the fit procedure are determined by taking the difference between the $\xicz\to\xizhz$ signal yield in the nominal fit and the signal yields in fits with the following modifications: 
(1) changing the order of polynomial for the smooth background, (2) floating the ratio of $n_{\rm broken}$ to $n_{\rm sig}$, (3) convolving the signal shapes of $\xicz\to\xiz\eta^{(\prime)}$ with a Gaussian function with a floating width, and (4) changing the broken-signal PDF smoothed by `rookeyspdf' to `roohistpdf'~\cite{rookeyspdf,roohistpdf}, corresponding to two algorithms for PDF estimation from simulated samples. 
The order of the polynomial for the background shape is common to the two experiments, and the corresponding uncertainty is extracted from a simultaneous fit for $\xicz$ signal yield in Belle and Belle II data.
Facing the worst signal-background ratio, the fitting uncertainties for the $\xicz\to\xiz\eta$ channel are larger than for the $\xicz\to\xizpiz$ and $\xicz\to\xizetap$ channels.
The total systematic uncertainty is obtained by adding the contributions from each source in quadrature. 

\subsection{Asymmetry parameter}

The sources of the systematic uncertainty on the asymmetry parameter measurement include the uncertainty on $\alpha(\xiz\to\Lambda\piz)$, the number of $\cos\theta_{\xiz}$ bins, and the uncertainties due to the fit procedure.
The relative uncertainty on $\alpha(\xiz\to\Lambda\piz)$=$-0.349\pm0.009$~\cite{pdg} is 2.6\%.
We change the number of $\cos\theta_{\xiz}$ bins from 5 to 4 and 6, and the largest difference in the extracted asymmetry parameter, 0.14, is taken as the associated systematic uncertainty.
The uncertainty from the fit procedure, 0.18, is determined using a similar procedure as for those in the branching fraction ratio measurements, where the width of the convolved Gaussian function is varied by $\pm$1$\sigma$ to obtain the uncertainty from reconstruction resolution.
We find that the systematic uncertainty due to the efficiency can be neglected since the efficiency is a multiplicative scale factor for the efficiency-corrected $\xicz$ signal yield in each $\cos\theta_{\xiz}$ bin and does not change the $\alpha$ value. 
As noted in section~\ref{datasets}, the simulated signal sample is weighted to match the observed value of $\alpha$.
When the weights are changed by the corresponding uncertainties ($\pm1\sigma$), the $\alpha$ measurement changes by less than 0.01: this effect is neglected.
We consider the migrations between adjacent $\cos\theta_{\xiz}$ bins in data due to the resolution effect by correcting signal yields with the matrix of migration rates obtained from MC simulations and find the difference in the extracted asymmetry parameter is smaller than 0.01 and can be neglected.
The measurement is insensitive to the $\Xi^0_c$ polarization, and no systematic uncertainty is included from this source~\cite{xic02xipiAlpha2001,xic02hypKstar2021}.
The systematic uncertainties from all sources are added in quadrature to obtain a value of 0.23.

\section{Summary and discussion}

We report the first measurements on $\xicz\to\xizhz$ decays, using the combined Belle and Belle~II data samples corresponding to a total integrated luminosity of about 1.4~$\mathrm{ab}^{-1}$.
The branching fractions of $\xicz\to\xizhz$ relative to $\BR(\xicz\to\xipi)$ are measured to be
\begin{equation} \label{rbf1}\BR(\xicz\to\xizpiz)/\BR(\xicz\to\xipi) = 0.48\pm 0.02 ({\rm stat}) \pm 0.03 ({\rm syst}),\end{equation}
\begin{equation} \label{rbf2}\BR(\xicz\to\xizeta)/\BR(\xicz\to\xipi) = 0.11\pm 0.01 ({\rm stat}) \pm 0.01 ({\rm syst}),\end{equation}
and
\begin{equation} \label{rbf3}\BR(\xicz\to\xizetap)/\BR(\xicz\to\xipi) = 0.08\pm 0.02 ({\rm stat}) \pm 0.01 ({\rm syst}),\end{equation}
where the first uncertainties are statistical and the second are systematic.
Taking $\BR(\xicz\to\xipi)=(1.43\pm0.27)\%$~\cite{pdg}, the absolute branching fractions are measured to be
\begin{equation}\BR(\xicz\to\xizpiz) = (6.9 \pm 0.3 ({\rm stat}) \pm 0.5 ({\rm syst}) \pm 1.3 ({\rm norm})) \times 10^{-3},\end{equation}
\begin{equation}\BR(\xicz\to\xizeta) = (1.6 \pm 0.2 ({\rm stat}) \pm 0.2 ({\rm syst}) \pm 0.3 ({\rm norm})) \times 10^{-3},\end{equation}
and
\begin{equation}\BR(\xicz\to\xizetap) = (1.2 \pm 0.3 ({\rm stat}) \pm 0.1 ({\rm syst}) \pm 0.2 ({\rm norm})) \times 10^{-3},\end{equation}
where the third uncertainty is from $\BR(\xicz\to\xipi)$.
We measure the asymmetry parameter
\begin{equation}\alpha(\xicz\to\xizpiz) = -0.90\pm0.15({\rm stat})\pm0.23({\rm syst})\end{equation}
for the first time.
Due to the limited data sample size, the asymmetry parameters for $\xicz\to\xizeta$ and $\xicz\to\xizetap$ are not measured, but will become accessible with the larger data samples to be collected by Belle~II in the future.

\begin{figure}[htbp]	 
	\centering
	\includegraphics[width=15cm]{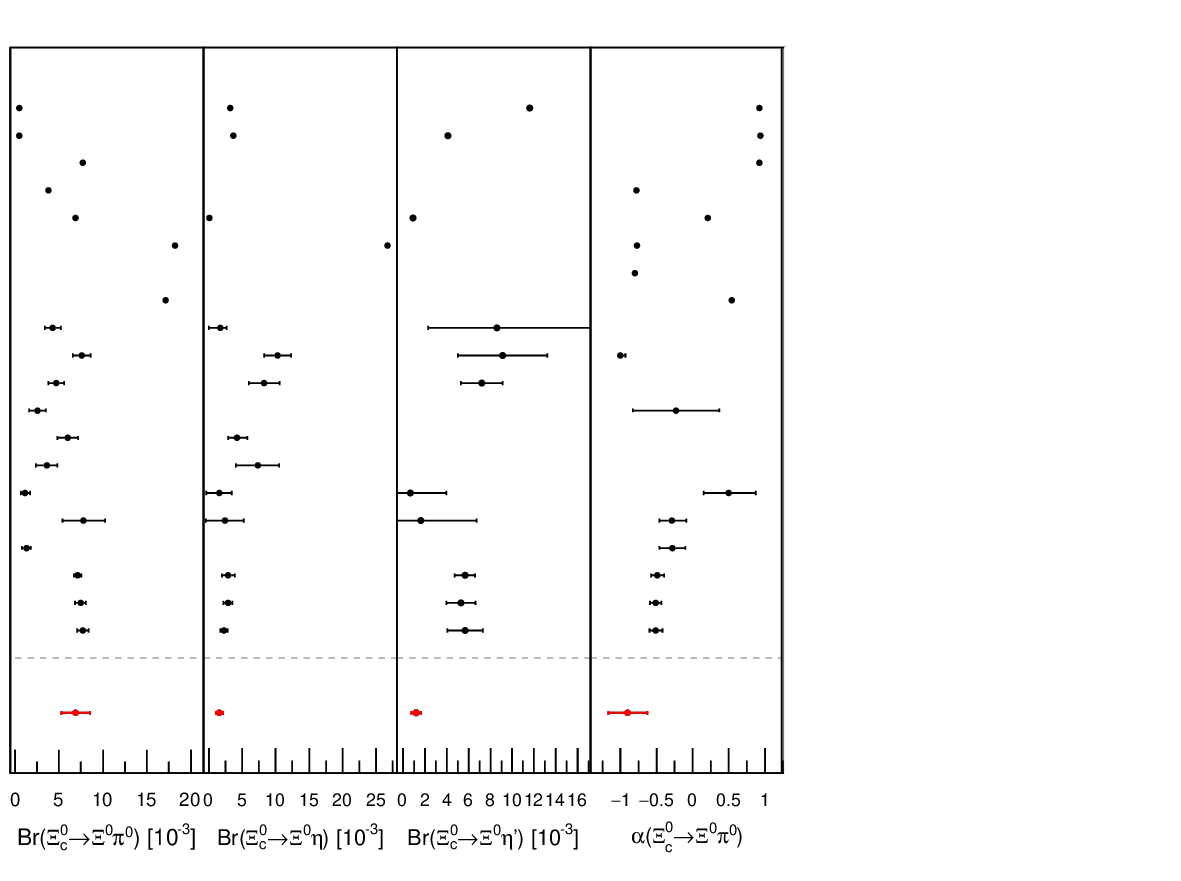}	
\put(-420,290){\bf (a)}\put(-350,290){\bf (b)} \put(-280,290){\bf (c)} \put(-210,290){\bf (d)}
{\scriptsize 
	\put(-140,280){K$\rm\ddot{o}$rner, Kr$\rm\ddot{a}$mer~\cite{theory1quark1992}, Quark} 
	\put(-140,270){Ivanov~\etal~\cite{theory5quark1998}, Quark} 
	\put(-140,260){Xu, Kamal~\cite{theory2pole1992}, Pole} 
	\put(-140,250){Cheng, Tseng~\cite{theory3poleca1993}, Pole} 
	\put(-140,240){{\.Z}enczykowski~\cite{theory4poleca1994}, Pole}
	\put(-140,230){Zou~\etal~\cite{theory10poleca2020}, Pole} 
	\put(-140,220){Sharma, Verma~\cite{theory6ca1999}, CA} 
	\put(-140,210){Cheng, Tseng~\cite{theory3poleca1993}, CA}
	\put(-140,200){Geng~\etal~\cite{theory7su3f2018}, $\rm SU(3)_F$} 
	\put(-140,190){Geng~\etal~\cite{theory8su3f2019}, $\rm SU(3)_F$}
	\put(-140,180){Zhao~\etal~\cite{theory9su3f2020}, $\rm SU(3)_F$} 
	\put(-140,170){Huang~\etal~\cite{theory11su3f2022}, $\rm SU(3)_F$} 
	\put(-140,160){Hsiao~\etal~\cite{theory12su3f2022}, $\rm SU(3)_F$} 
	\put(-140,150){Hsiao~\etal~\cite{theory12su3f2022}, $\rm SU(3)_F$-breaking}
	\put(-140,140){Zhong~\etal~\cite{theory13su3f2023}, $\rm SU(3)_F$} 
	\put(-140,130){Zhong~\etal~\cite{theory13su3f2023}. $\rm SU(3)_F$-breaking}  
	\put(-140,120){Xing~\etal~\cite{theory14su3f2023}, $\rm SU(3)_F$}  
	\put(-140,110){Geng~\etal~\cite{theory15su3f2024}, $\rm SU(3)_F$} 
	\put(-140,100){Zhong~\etal~\cite{theory16su3f2024}, Diagrammatic-$\rm SU(3)_F$}  
	\put(-140,90){Zhong~\etal~\cite{theory16su3f2024}, Irreducible-$\rm SU(3)_F$}  
	\put(-140,60){\color{red}{This measurement}} } 
\caption{The comparisons of the measured (a) $\BR(\xicz\to\xizpiz)$, (b) $\BR(\xicz\to\xizeta)$, (c) $\BR(\xicz\to\xizetap)$, and (d) $\alpha(\xicz\to\xizpiz)$ with theoretical predictions~\cite{theory1quark1992,theory5quark1998,theory2pole1992,theory3poleca1993,theory4poleca1994,theory10poleca2020,theory6ca1999,theory7su3f2018,theory8su3f2019,theory9su3f2020,theory11su3f2022,theory12su3f2022,theory13su3f2023,theory14su3f2023,theory15su3f2024,theory16su3f2024}, corresponding to the values in table~1.
	The dots and error bars show the center values and uncertainties, respectively, where the dots without error bars mean that no theoretical uncertainty is available.}
	\label{comparsion}
\end{figure}

Figure~\ref{comparsion} shows the comparisons of our measurements with theoretical predictions from table~\ref{theoryPredict}.
A recent result~\cite{theory13su3f2023} based on the $\rm SU(3)_F$-breaking model is consistent with each measured $\BR(\xicz\to\xiz\hz)$.
The measured value of $\alpha(\xicz\to\xiz\piz)$ is consistent with predictions based on the pole model~\cite{theory3poleca1993,theory10poleca2020}, CA~\cite{theory6ca1999}, and $\rm SU(3)_F$ flavor symmetry~\cite{theory8su3f2019} approaches. 
The central values of our measurements of the absolute branching fractions and asymmetry parameter of $\Xi_c^0 \to \Xi^0\pi^0$, indicate that the covariant confined quark model~\cite{theory1quark1992,theory5quark1998} is mildly disfavored for each result,
and disagree with the predictions by more than $2\sigma$ for the following: 
(1) $\BR(\xicz\to\xiz\piz)$ in refs.~\cite{theory3poleca1993,theory10poleca2020,theory11su3f2022,theory14su3f2023};
(2) $\BR(\xicz\to\xiz\eta)$ in refs.~\cite{theory4poleca1994,theory10poleca2020,theory8su3f2019,theory15su3f2024,theory16su3f2024};
(3) $\BR(\xicz\to\xiz\etap)$ in refs.~\cite{theory9su3f2020,theory15su3f2024,theory16su3f2024};
and (4) $\alpha(\xicz\to\xiz\piz)$ in refs.~\cite{theory2pole1992,theory4poleca1994,theory13su3f2023,theory14su3f2023,theory15su3f2024,theory16su3f2024}.
The results for the ratios, (\ref{rbf1}), (\ref{rbf2}), and (\ref{rbf3}), are independent of the $\xicz$ absolute branching fraction scale and may also be compared to theoretical models.

\acknowledgments
This work, based on data collected using the Belle II detector, which was built and commissioned prior to March 2019,
was supported by
Higher Education and Science Committee of the Republic of Armenia Grant No.~23LCG-1C011;
Australian Research Council and Research Grants
No.~DP200101792, 
No.~DP210101900, 
No.~DP210102831, 
No.~DE220100462, 
No.~LE210100098, 
and
No.~LE230100085; 
Austrian Federal Ministry of Education, Science and Research,
Austrian Science Fund
No.~P~34529,
No.~J~4731,
No.~J~4625,
and
No.~M~3153,
and
Horizon 2020 ERC Starting Grant No.~947006 ``InterLeptons'';
Natural Sciences and Engineering Research Council of Canada, Compute Canada and CANARIE;
National Key R\&D Program of China under Contract No.~2022YFA1601903,
National Natural Science Foundation of China and Research Grants
No.~11575017,
No.~11761141009,
No.~11705209,
No.~11975076,
No.~12135005,
No.~12150004,
No.~12161141008,
and
No.~12175041,
and Shandong Provincial Natural Science Foundation Project~ZR2022JQ02;
the Czech Science Foundation Grant No.~22-18469S 
and
Charles University Grant Agency project No.~246122;
European Research Council, Seventh Framework PIEF-GA-2013-622527,
Horizon 2020 ERC-Advanced Grants No.~267104 and No.~884719,
Horizon 2020 ERC-Consolidator Grant No.~819127,
Horizon 2020 Marie Sklodowska-Curie Grant Agreement No.~700525 ``NIOBE''
and
No.~101026516,
and
Horizon 2020 Marie Sklodowska-Curie RISE project JENNIFER2 Grant Agreement No.~822070 (European grants);
L'Institut National de Physique Nucl\'{e}aire et de Physique des Particules (IN2P3) du CNRS
and
L'Agence Nationale de la Recherche (ANR) under grant ANR-21-CE31-0009 (France);
BMBF, DFG, HGF, MPG, and AvH Foundation (Germany);
Department of Atomic Energy under Project Identification No.~RTI 4002,
Department of Science and Technology,
and
UPES SEED funding programs
No.~UPES/R\&D-SEED-INFRA/17052023/01 and
No.~UPES/R\&D-SOE/20062022/06 (India);
Israel Science Foundation Grant No.~2476/17,
U.S.-Israel Binational Science Foundation Grant No.~2016113, and
Israel Ministry of Science Grant No.~3-16543;
Istituto Nazionale di Fisica Nucleare and the Research Grants BELLE2;
Japan Society for the Promotion of Science, Grant-in-Aid for Scientific Research Grants
No.~16H03968,
No.~16H03993,
No.~16H06492,
No.~16K05323,
No.~17H01133,
No.~17H05405,
No.~18K03621,
No.~18H03710,
No.~18H05226,
No.~19H00682, 
No.~20H05850,
No.~20H05858,
No.~22H00144,
No.~22K14056,
No.~22K21347,
No.~23H05433,
No.~26220706,
and
No.~26400255,
and
the Ministry of Education, Culture, Sports, Science, and Technology (MEXT) of Japan;  
National Research Foundation (NRF) of Korea Grants
No.~2016R1\-D1A1B\-02012900,
No.~2018R1\-A2B\-3003643,
No.~2018R1\-A6A1A\-06024970,
No.~2019R1\-I1A3A\-01058933,
No.~2021R1\-A6A1A\-03043957,
No.~2021R1\-F1A\-1060423,
No.~2021R1\-F1A\-1064008,
No.~2022R1\-A2C\-1003993,
and
No.~RS-2022-00197659,
Radiation Science Research Institute,
Foreign Large-Size Research Facility Application Supporting project,
the Global Science Experimental Data Hub Center of the Korea Institute of Science and Technology Information
and
KREONET/GLORIAD;
Universiti Malaya RU grant, Akademi Sains Malaysia, and Ministry of Education Malaysia;
Frontiers of Science Program Contracts
No.~FOINS-296,
No.~CB-221329,
No.~CB-236394,
No.~CB-254409,
and
No.~CB-180023, and SEP-CINVESTAV Research Grant No.~237 (Mexico);
the Polish Ministry of Science and Higher Education and the National Science Center;
the Ministry of Science and Higher Education of the Russian Federation
and
the HSE University Basic Research Program, Moscow;
University of Tabuk Research Grants
No.~S-0256-1438 and No.~S-0280-1439 (Saudi Arabia);
Slovenian Research Agency and Research Grants
No.~J1-9124
and
No.~P1-0135;
Agencia Estatal de Investigacion, Spain
Grant No.~RYC2020-029875-I
and
Generalitat Valenciana, Spain
Grant No.~CIDEGENT/2018/020;
The Knut and Alice Wallenberg Foundation (Sweden), Contracts No.~2021.0174 and No.~2021.0299;
National Science and Technology Council,
and
Ministry of Education (Taiwan);
Thailand Center of Excellence in Physics;
TUBITAK ULAKBIM (Turkey);
National Research Foundation of Ukraine, Project No.~2020.02/0257,
and
Ministry of Education and Science of Ukraine;
the U.S. National Science Foundation and Research Grants
No.~PHY-1913789 
and
No.~PHY-2111604, 
and the U.S. Department of Energy and Research Awards
No.~DE-AC06-76RLO1830, 
No.~DE-SC0007983, 
No.~DE-SC0009824, 
No.~DE-SC0009973, 
No.~DE-SC0010007, 
No.~DE-SC0010073, 
No.~DE-SC0010118, 
No.~DE-SC0010504, 
No.~DE-SC0011784, 
No.~DE-SC0012704, 
No.~DE-SC0019230, 
No.~DE-SC0021274, 
No.~DE-SC0021616, 
No.~DE-SC0022350, 
No.~DE-SC0023470; 
and
the Vietnam Academy of Science and Technology (VAST) under Grants
No.~NVCC.05.12/22-23
and
No.~DL0000.02/24-25.

These acknowledgements are not to be interpreted as an endorsement of any statement made
by any of our institutes, funding agencies, governments, or their representatives.

We thank the SuperKEKB team for delivering high-luminosity collisions;
the KEK cryogenics group for the efficient operation of the detector solenoid magnet;
the KEK Computer Research Center for on-site computing support; the NII for SINET6 network support;
and the raw-data centers hosted by BNL, DESY, GridKa, IN2P3, INFN, 
and the University of Victoria.

\renewcommand{\baselinestretch}{1.2}

\begin{appendices}
\appendix
\section{$M(\xiz\piz)$ spectra in $\cos\theta_{\xiz}$ bins \label{alphaFits}}
\begin{figure}[htbp]	
	\centering
	\includegraphics[width=6.0cm]{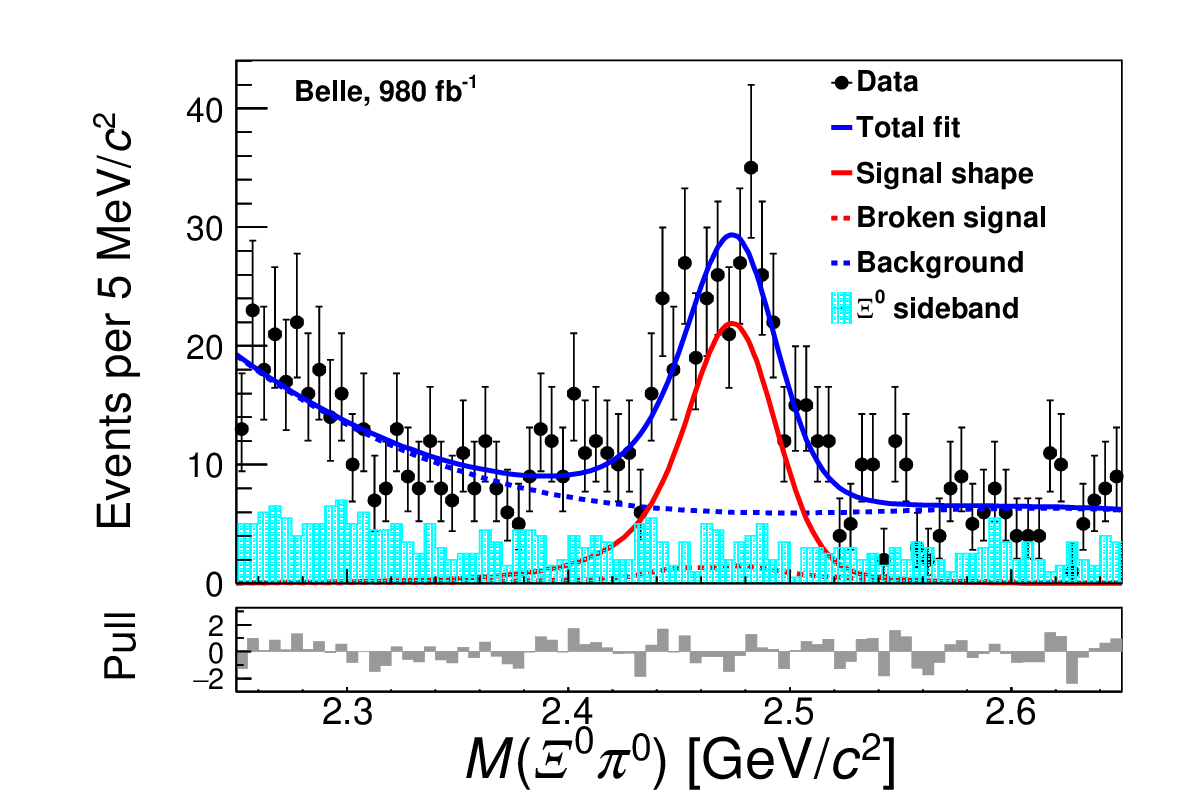}	\put(-180,90){\bf (a)}		
	\includegraphics[width=6.0cm]{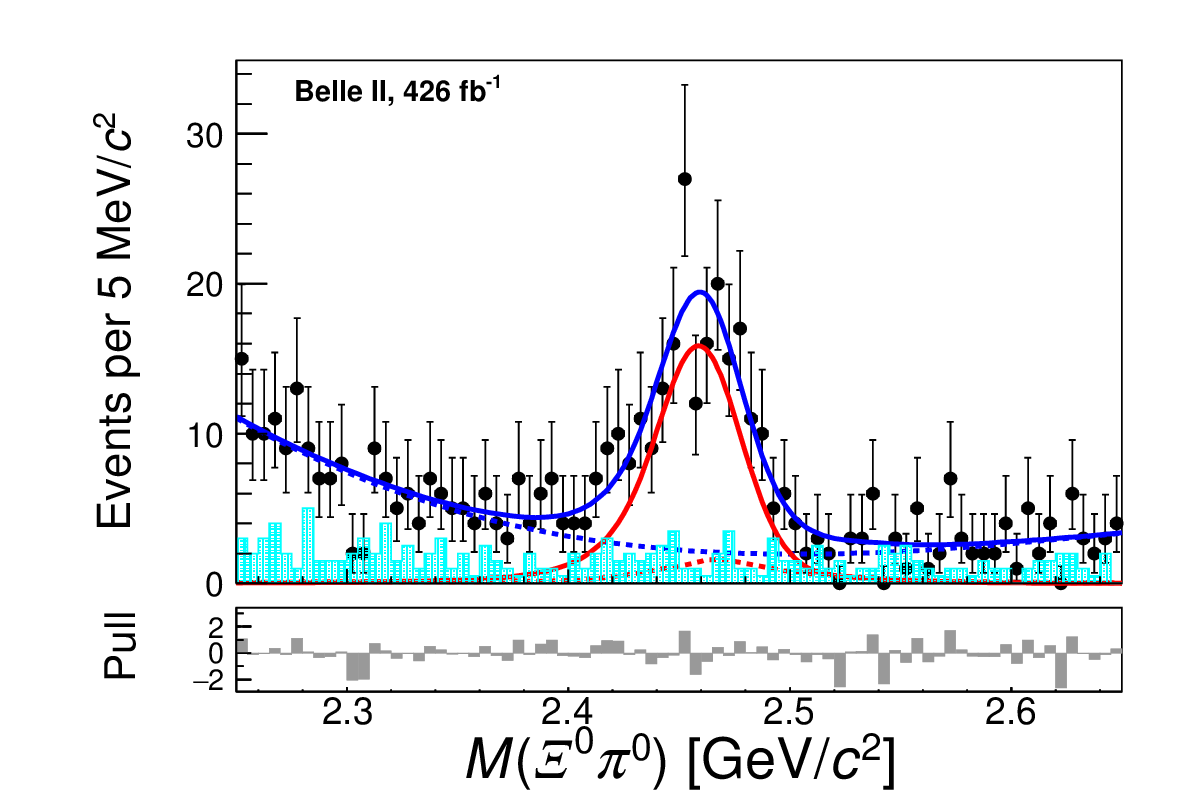}	
	
	\includegraphics[width=6.0cm]{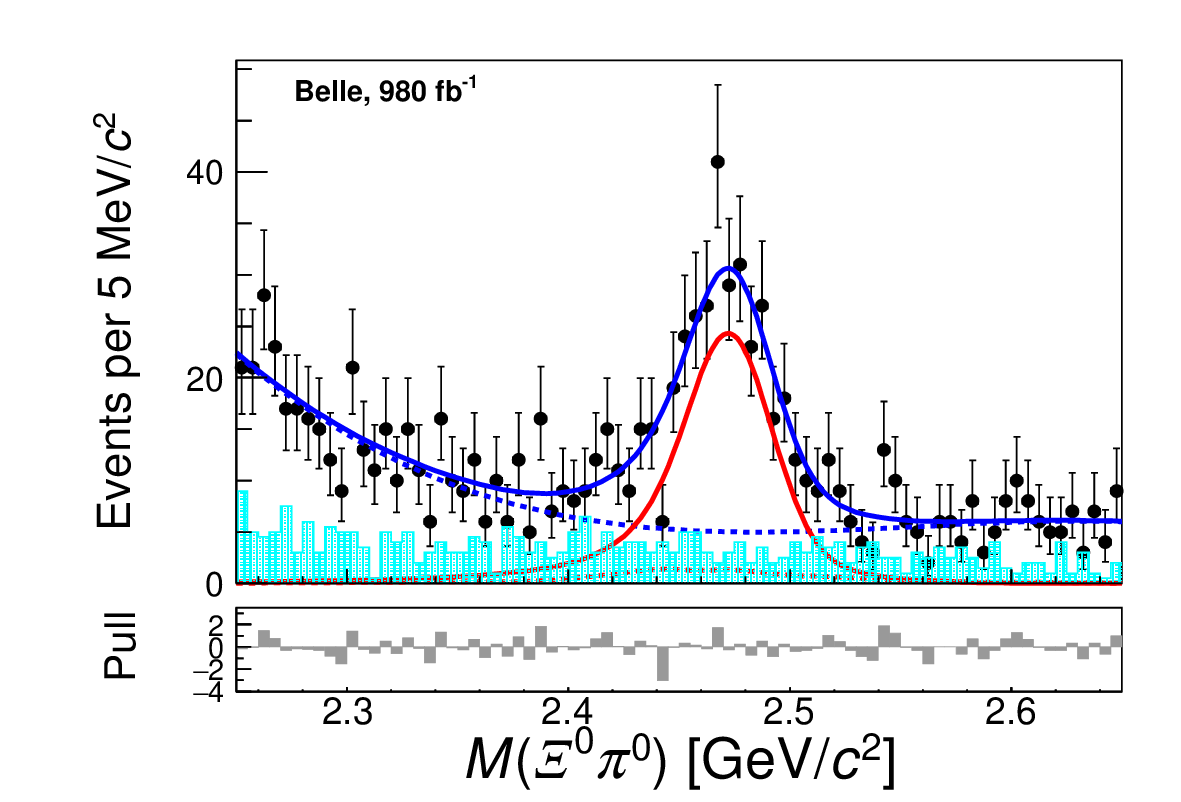}	\put(-180,90){\bf (b)}		
	\includegraphics[width=6.0cm]{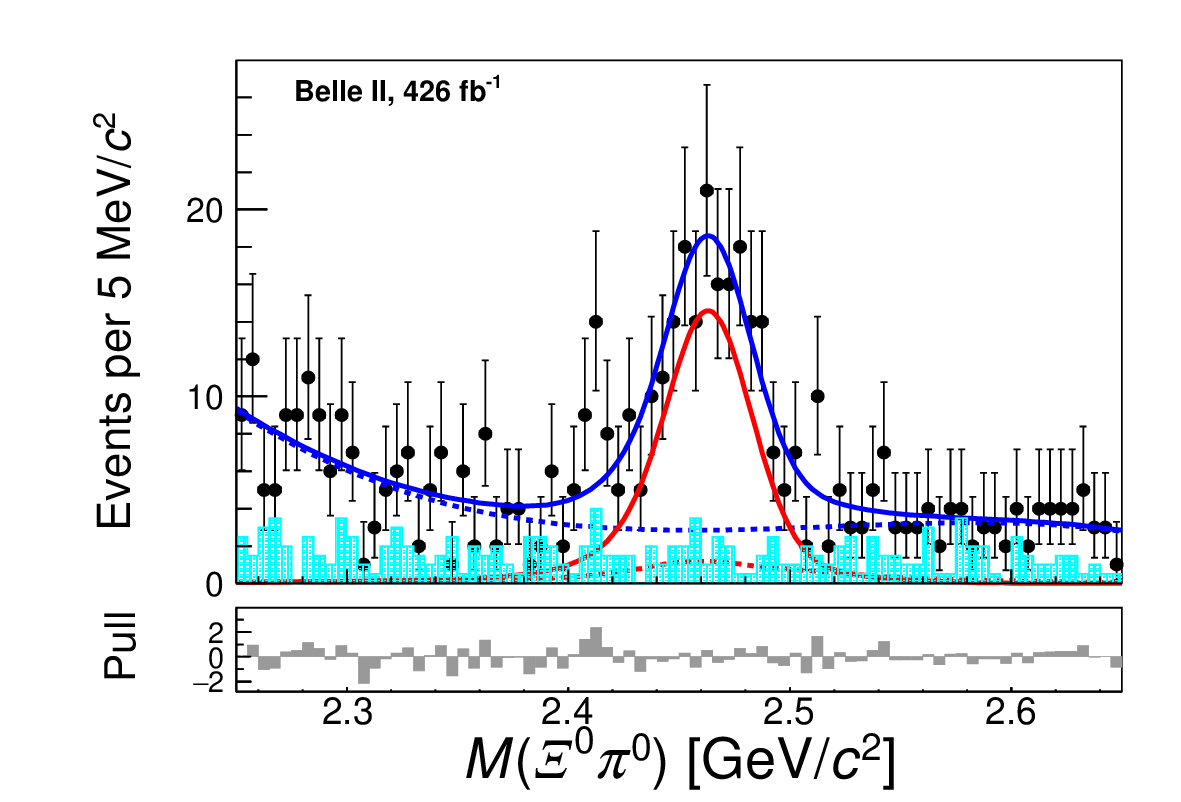}	
	
	\includegraphics[width=6.0cm]{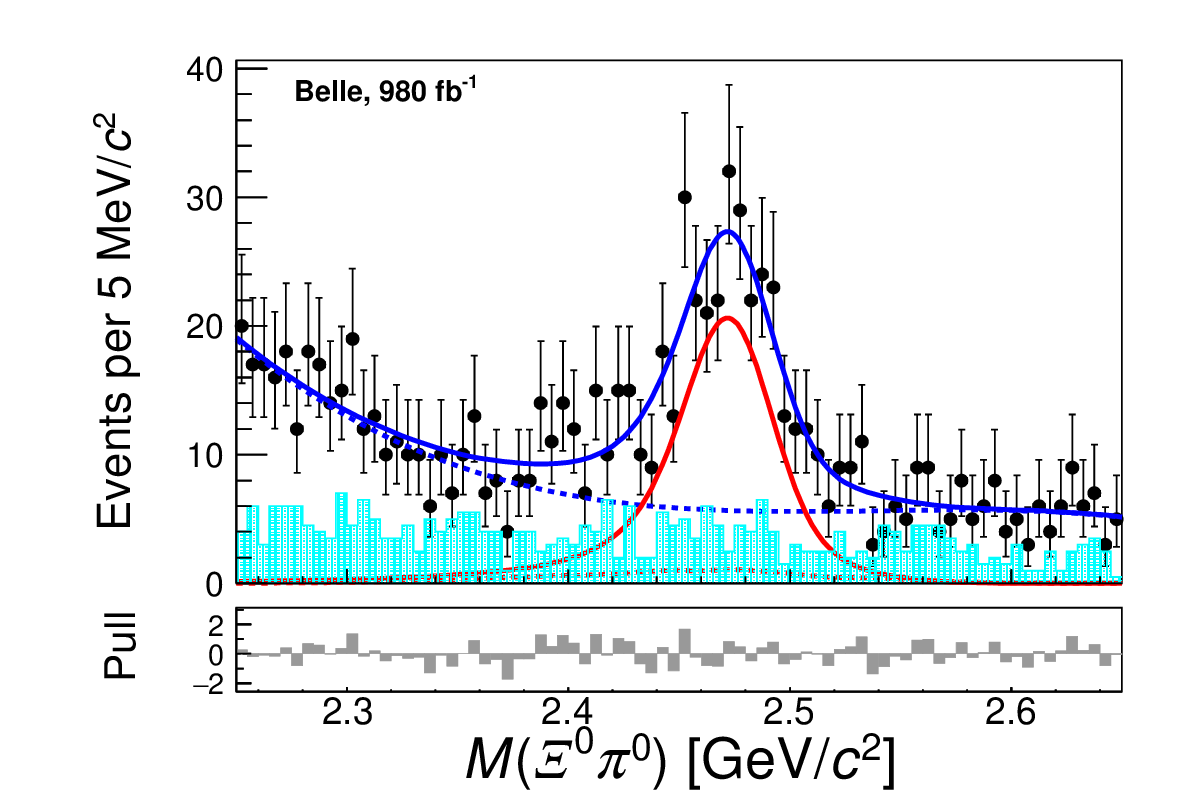}	\put(-180,90){\bf (c)}		
	\includegraphics[width=6.0cm]{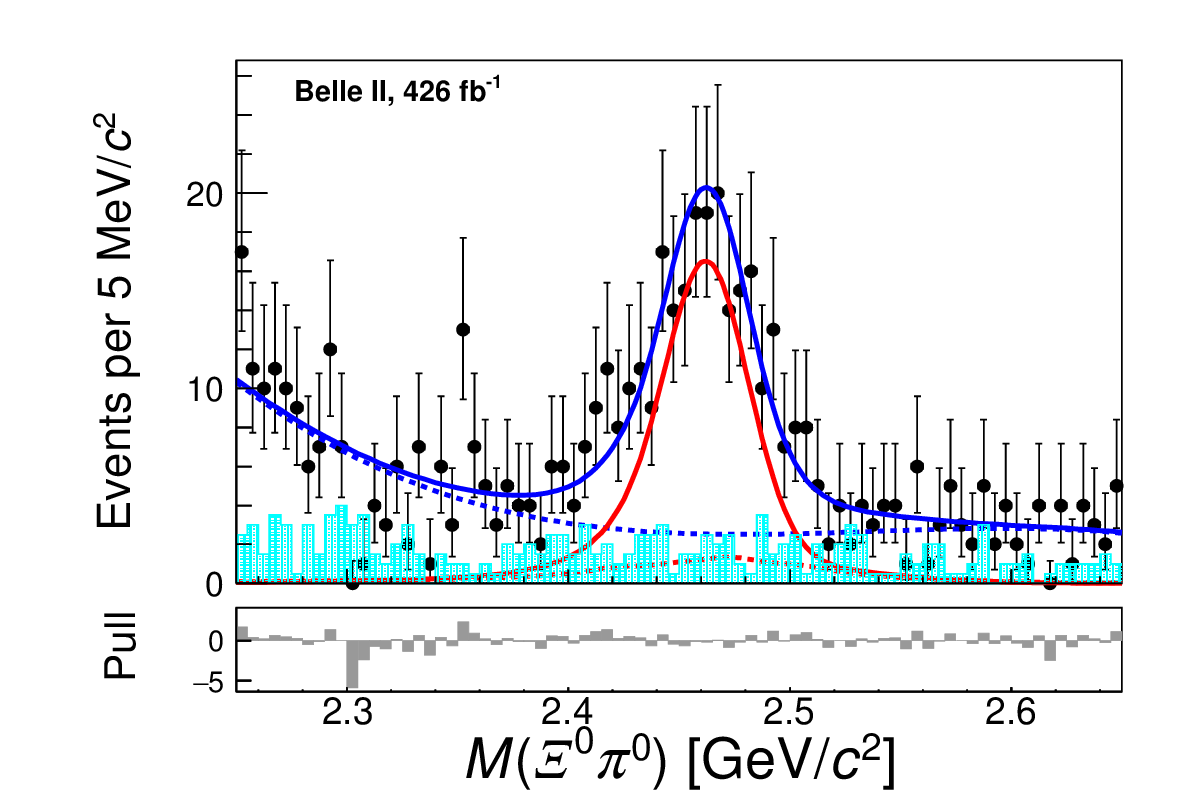}	
	
	\includegraphics[width=6.0cm]{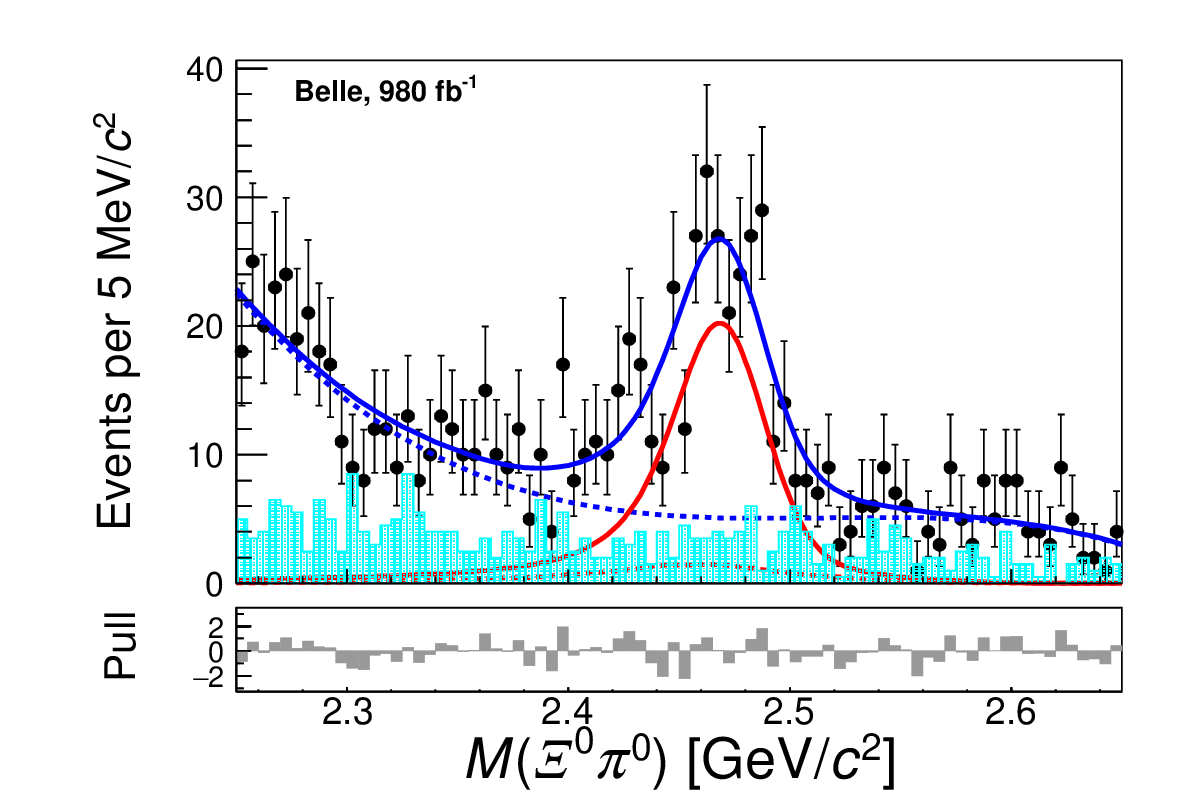}	\put(-180,90){\bf (d)}		
	\includegraphics[width=6.0cm]{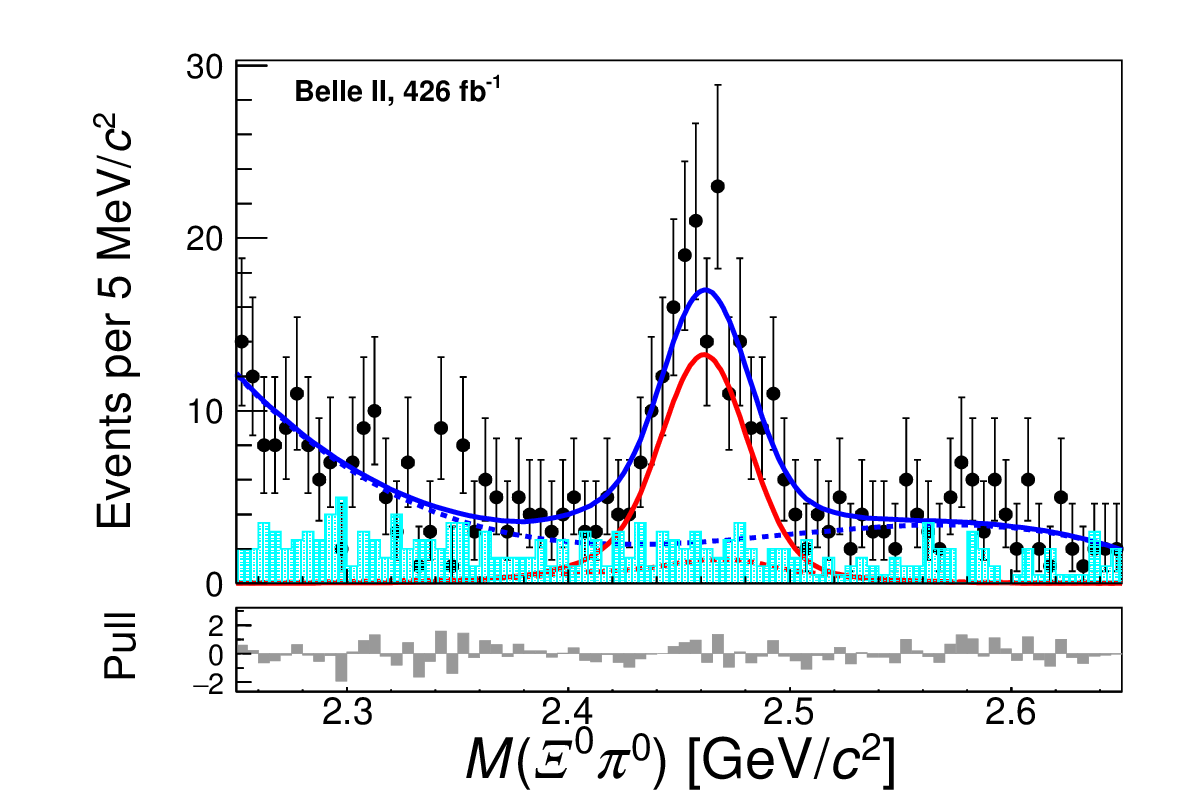}	
	
	\includegraphics[width=6.0cm]{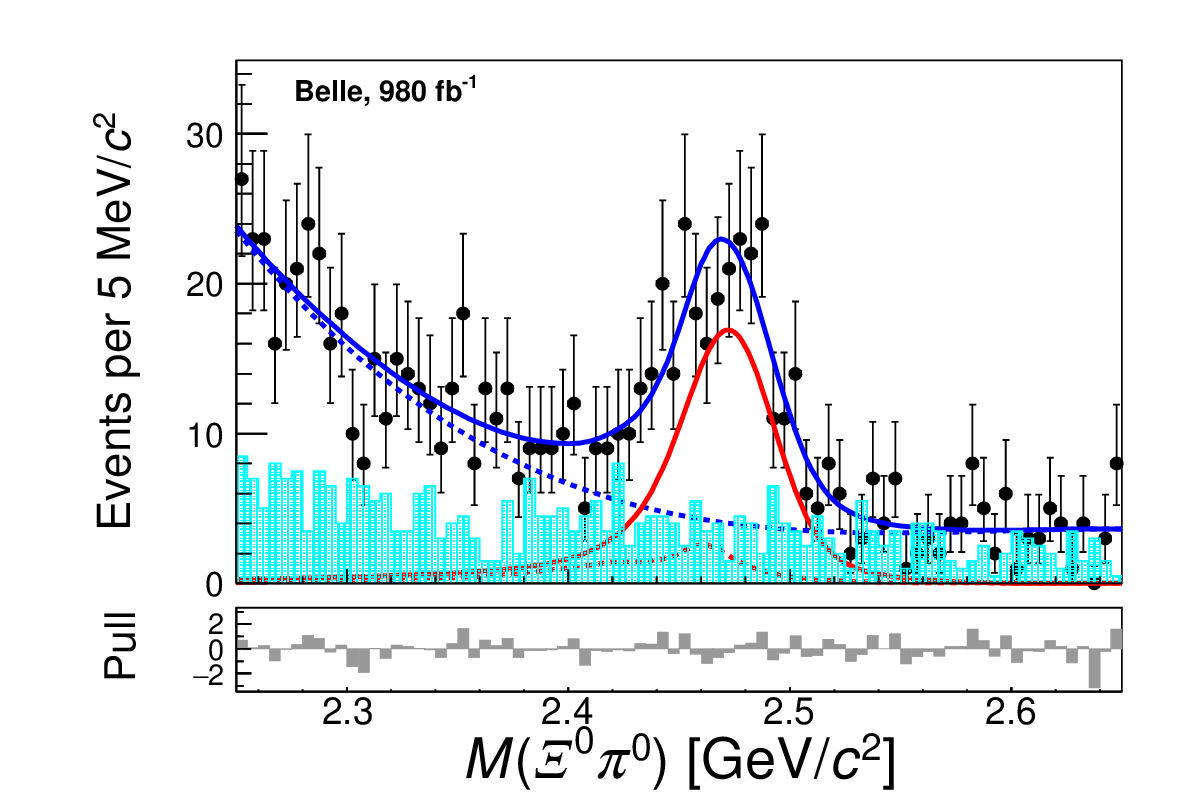}	\put(-180,90){\bf (e)}		
	\includegraphics[width=6.0cm]{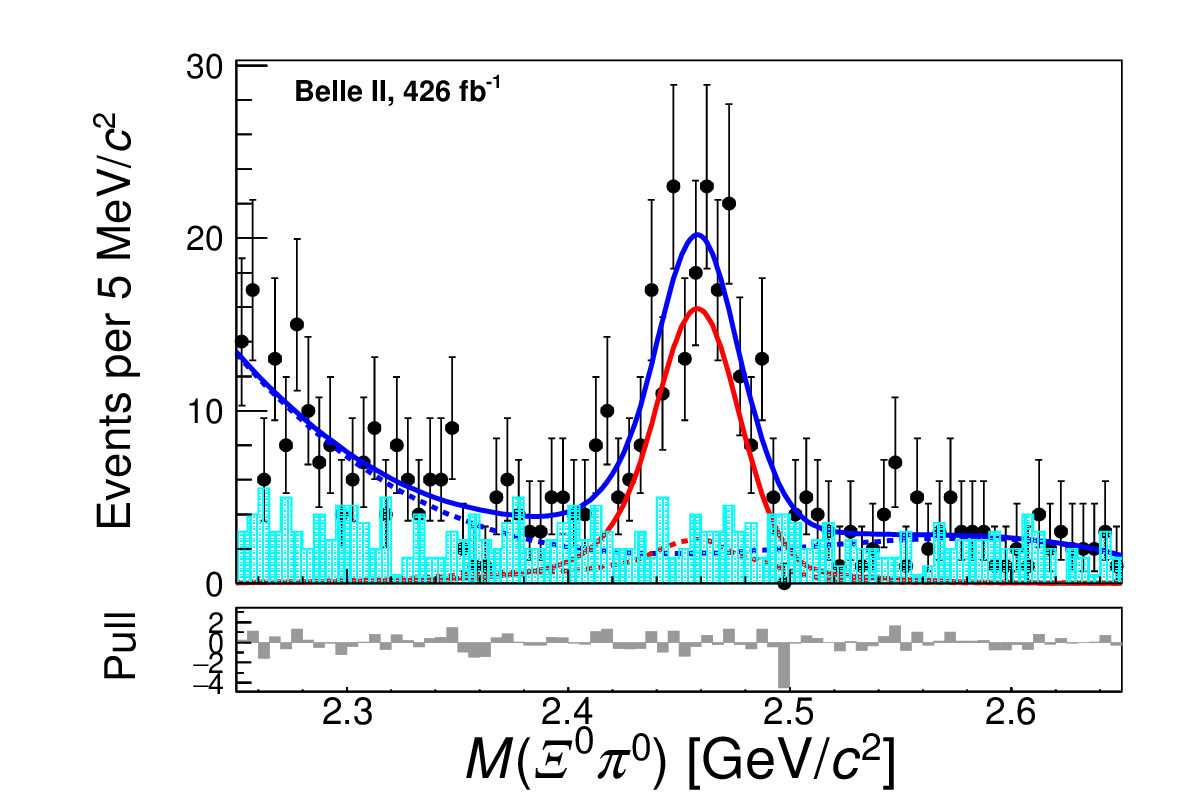}

	\caption{Invariant mass distributions of $\xicz\to\xizpiz$ candidates from (left) Belle and (right) Belle II data samples in $\cos\theta_{\xiz}$ bins of (a) $(-1.0, -0.6)$, (b) $(-0.6, -0.2)$, (c) $(-0.2, 0.2)$, (d) $(0.2, 0.6)$ and (e) $(0.6, 1.0)$.
		The markers with error bars represent the data.
		The solid blue curves, solid red curves, dashed red curves, and dashed blue curves show the total fit, signal shape, broken-signal shape, and smooth backgrounds, respectively.
		The cyan histograms show the data from the $\xiz$ mass sidebands.
	}
	\label{Mxic02xi0h0DataAlpha}
\end{figure}

Distributions of $M(\xiz\piz)$ in bins of $\cos\theta_{\xiz}$ are shown in figure~\ref{Mxic02xi0h0DataAlpha} with fit results overlaid.

\end{appendices}


\begin{thebibliography}{**}
\addtolength{\itemsep}{-0.3 ex}
\bibitem{xic0absbf2019} Belle Collaboration, {\em First Measurements of Absolute Branching Fractions of the ${\mathrm{\ensuremath{\Xi}}}_{c}^{0}$ Baryon at Belle}, Phys. Rev. Lett. {\bf 122} (2019) 082001.

\bibitem{xic02hypKstar2021} Belle Collaboration, {\em Measurements of branching fractions and asymmetry parameters of $ {\Xi}_c^0\to \Lambda {\overline{K}}^{\ast 0} $, $ {\Xi}_c^0\to {\Sigma}^0{\overline{K}}^{\ast 0} $, and $ {\Xi}_c^0\to {\Sigma}^{+}{K}^{\ast -}$ decays at Belle}, JHEP {\bf 06} (2021) 160.

\bibitem{xic02hypK2022} Belle Collaboration, {\em  {Measurements of the branching fractions of ${\mathrm{\ensuremath{\Xi}}}_{c}^{0}\ensuremath{\rightarrow}\mathrm{\ensuremath{\Lambda}}{K}_{S}^{0}$, ${\mathrm{\ensuremath{\Xi}}}_{c}^{0}\ensuremath{\rightarrow}{\mathrm{\ensuremath{\Sigma}}}^{0}{K}_{S}^{0}$, and ${\mathrm{\ensuremath{\Xi}}}_{c}^{0}\ensuremath{\rightarrow}{\mathrm{\ensuremath{\Sigma}}}^{+}{K}^{\ensuremath{-}}$ decays at Belle}}, Phys. Rev. D {\bf 105} (2022) L011102.
\bibitem{charmedBaryon2022} Hai-Yang Cheng, {\em Charmed baryon physics circa 2021}, Chin. J. Phys. {\bf 78} (2022) 324-362. 

\bibitem{theory1quark1992} J. G. K$\rm\ddot{o}$rner and M. Kr$\rm\ddot{a}$mer, {\em Exclusive non-leptonic charm baryon decays}, Z. Phys. C {\bf 55} (1992) 659.
\bibitem{theory5quark1998} M. A. Ivanov, J. G. Korner, V. E. Lyubovitskij, and A. G. Rusetsky, {\em Exclusive nonleptonic decays of bottom and charm baryons in a relativistic three-quark model: Evaluation of nonfactorizing diagrams}, Phys. Rev. D {\bf 57} (1998) 5632.
\bibitem{theory2pole1992} Q. P. Xu and A. N. Kamal, {\em Cabibbo-favored nonleptonic decays of charmed baryons}, Phys. Rev. D {\bf 46} (1992) 270.
\bibitem{theory3poleca1993} H. Y. Cheng and B. Tseng, {\em Cabibbo-allowed nonleptonic weak decays of charmed baryons}, Phys. Rev. D {\bf 48} (1993) 4188.
\bibitem{theory4poleca1994} P. {\.Z}enczykowski, {\em Nonleptonic charmed-baryon decays: Symmetry properties of parity-violating amplitudes}, Phys. Rev. D {\bf 50} (1994) 5787.
\bibitem{theory10poleca2020} J. Q. Zou, F. R. Xu, G. B. Meng, and H. Y. Cheng, {\em Two-body hadronic weak decays of antitriplet charmed baryons}, Phys. Rev. D {\bf 101} (2020) 014011.
\bibitem{theory6ca1999} K. K. Sharma and R. C. Verma, {\em  A study of weak mesonic decays of $\Lambda_c$ and $\Xi_C$ baryons on the basis of HQET results}, Eur. Phys. J. C {\bf 7} (1999) 217.
\bibitem{theory7su3f2018} C. Q. Geng, Y. K. Hsiao, C. W. Liu, and T. H. Tsai, {\em Antitriplet charmed baryon decays with SU(3) flavor symmetry}, Phys. Rev. D {\bf 97} (2018) 073006.
\bibitem{theory8su3f2019} C. Q. Geng, C. W. Liu, and T. H. Tsai, {\em Asymmetries of anti-triplet charmed baryon decays}, Phys. Lett. B {\bf 794} (2019) 19. 
\bibitem{theory9su3f2020} H. J. Zhao, Y. L. Wang, Y. K. Hsiao, and Y. Yu, {\em A Diagrammatic Analysis of Two-Body Charmed Baryon Decays with Flavor Symmetry}, JHEP {\bf 02} (2020) 165.
\bibitem{theory11su3f2022} F. Huang, Z. P. Xing, and X. Z. He, {\em A global analysis of charmless two body hadronic decays for anti-triplet charmed baryons}, JHEP {\bf 03} (2022) 143.
\bibitem{theory12su3f2022} Y. K. Hsiao, Y. L. Wang, and H. J. Zhao, {\em Equivalent $\rm SU(3)_{f}$ approaches for two-body anti-triplet charmed baryon decays}, JHEP {\bf 09} (2022) 35.
\bibitem{theory13su3f2023} H. Zhong, F. Xu, Q. Wen, and Y. Gu, {\em Weak decays of antitriplet charmed baryons from the perspective of flavor symmetry}, JHEP {\bf 02} (2023) 235.
\bibitem{theory14su3f2023} Z. P. Xing, \etal, {\em Global analysis of measured and unmeasured hadronic two-body weak decays of antitriplet charmed baryons}, Phys. Rev. D {\bf 108} (2023) 053004.
\bibitem{theory15su3f2024} C. Q. Geng, \etal {\em Complete determination of $SU(3{)}_{F}$ amplitudes and strong phase in ${\mathrm{\ensuremath{\Lambda}}}_{c}^{+}\ensuremath{\rightarrow}{\mathrm{\ensuremath{\Xi}}}^{0}{K}^{+}$}, Phys. Rev. D {\bf 109} (2024) L071302. 
\bibitem{theory16su3f2024} H. Zhong, F. Xu, and H. Y. Cheng, {\em Analysis of Hadronic Weak Decays of Charmed Baryons in the Topological Diagrammatic Approach}, Phys. Rev. D {\bf 109} (2024) 114027.


\bibitem{alphaFunction} R. E. Behrends, {\em Photon Decay of Hyperons}, Phys. Rev. {\bf 111} (1958) 1691.


\bibitem{xic02xipiAlpha2001} CLEO Collaboration, {\em Measurement of the decay asymmetry parameters in $\xicz\to\xipi$}, Phys. Rev. D {\bf{63}} (2001) 111102.
\bibitem{xic0semilptANDalpha2021} Belle Collaboration, {\em Measurements of the Branching Fractions of the Semileptonic Decays ${\mathrm{\ensuremath{\Xi}}}_{c}^{0}\ensuremath{\rightarrow}{\mathrm{\ensuremath{\Xi}}}^{\ensuremath{-}}{\ensuremath{\ell}}^{+}{\ensuremath{\nu}}_{\ensuremath{\ell}}$ and the Asymmetry Parameter of ${\mathrm{\ensuremath{\Xi}}}_{c}^{0}\ensuremath{\rightarrow}{\mathrm{\ensuremath{\Xi}}}^{\ensuremath{-}}{\ensuremath{\pi}}^{+}$}, Phys. Rev. Lett. {\bf 127} (2021) 121803.





\bibitem{Belle1} Belle Collaboration, {\em The Belle detector}, Nucl. Instr. and Methods Phys. Res. Sect. A {\bf 479} (2002) 117. 
\bibitem{Belle2} Belle Collaboration, {\em Physics achievements from the Belle experiment}, Prog. Theor. Exp. Phys. {\bf 2012} (2012) 04D001.
\bibitem{KEKB1} S. Kurokawa and E. Kikutani, {\em Overview of the KEKB accelerators}, Nucl. Instr. and Methods Phys. Res. Sect. A {\bf 499} (2003) 1, and other papers included in this volume.
\bibitem{KEKB2} T. Abe {\em et al.}, {\em Achievements of KEKB}, Prog. Theor. Exp. Phys. {\bf 2013} (2013) 03A001, and references therein.

\bibitem{BelleII} Belle II Collaboration, {\em Belle II Technical Design Report}, arXiv:1011.0352.
\bibitem{superKEKB} SuperKEKB Collaboration, {\em SuperKEKB Collider}, Nucl. Instr. and Methods Phys. Res. Sect. A {\bf 907}  (2018) 188. 



\bibitem{pythia1} T. Sj$\rm \ddot{o}$strand {\em et al.}, {\em High-energy physics event generation with PYTHIA 6.1},  Comput. Phys. Commun. {\bf 135} (2001) 238.
\bibitem{pythia2} T. Sj$\rm \ddot{o}$strand {\em et al.}, {\em An introduction to PYTHIA 8.2}, Comput. Phys. Commun. {\bf 191} (2015) 159.
\bibitem{evtgen} D.J. Lange, {\em The EvtGen particle decay simulation package}, Nucl. Instr. and Methods Phys. Res. Sect. A {\bf 462} (2001) 152.
\bibitem{kkmc} S. Jadach, B. F. L. Ward and Z. W\c as, {\em The precision Monte Carlo event generator $\mathcal K \mathcal K$ for two-fermion final states in $e^+e^-$ collisions}, Comput. Phys. Commun. {\bf 130} (2000) 260.
\bibitem{photos} E. Barberio and Z. W{\c a}s, {\em PHOTOS: A Universal Monte Carlo for QED radiative corrections. Version 2.0}, Comput. Phys. Commun. {\bf 79} (1994) 291.
\bibitem{geant3} R. Brun {\em et al.}, {\em GEANT3}, CERN Report No. DD/EE/84-1 (1984).
\bibitem{geant4} GEANT4 collaboration, {\em GEANT4--a simulation toolkit},  Nucl. Instrum. Methods Phys. Sect. A {\bf 506} (2003) 250.


\bibitem{basf2} Belle II Framework Software Group, {\em The Belle II Core Software}, Comput. Softw. Big Sci. {\bf 3} (2019) 1.
\bibitem{b2bii} M.~Gelb {\it et al.}, {\em B2BII: Data Conversion from Belle to Belle II}, Comput. Softw. Big Sci. {\bf 2} (2018) 9.
\bibitem{b2fit}  J.-F. Krohn {\it et al.}, {\em Global decay chain vertex fitting at Belle II}, Nucl. Instrum. Methods Phys. Res., Sect. A {\bf 976} (2020) 164269.

\bibitem{BellePID1} E.~Nakano, {\em Belle PID}, Nucl. Instrum. Methods Phys. Res., Sect. A {\bf 494} (2002) 402.

\bibitem{omega2018} Belle Collaboration, {\em Observation of an Excited $\Omega^-$ Baryon}, Phys. Rev. Lett. {\bf 121} (2018) 052003.
\bibitem{xic02xi0kk2021} Belle Collaboration, {\em Measurement of the resonant and nonresonant branching ratios in ${\mathrm{\ensuremath{\Xi}}}_{c}^{0}\ensuremath{\rightarrow}{\mathrm{\ensuremath{\Xi}}}^{0}{K}^{+}{K}^{\ensuremath{-}}$}, Phys. Rev. D {\bf 103} (2021) 112002.
\bibitem{xic02xi0ll2023} Belle Collaboration, {\em Search for the semileptonic decays $\Xi_c^0\to\Xi^0\ell^+\ell^-$ at Belle}, Phys. Rev. D {\bf 109} (2024) 052003.

\bibitem{pdg} Particle Data Group, {\em The Review of Particle Physics}, Phys. Rev. D {\bf 110}, 030001 (2024).

\bibitem{PunziFOM} G. Punzi, {\em Sensitivity of searches for new signals and its optimization}, eConf {\bf C030908} (2003) MODT002. arXiv:physics/0308063.


		\bibitem{cbfunction} J. E. Gaiser, {\em Charmonium Spectroscopy From Radiative Decays of the $J/\psi$ and $\psi^{\prime}$}, Ph. D. thesis, Stanford Linear Accelerator Center, Stanford University, Report No. SLAC-R-255,	1982.
		\bibitem{topoana} X. Y. Zhou, S. X. Du, G. Li, and C. P. Shen, {\em TopoAna: A generic tool for the event type analysis of inclusive Monte-Carlo samples in high energy physics experiments}, Comput. Phys. Commun. {\bf 258} (2021) 107540.
		\bibitem{rookeyspdf} K. S. Cranmer, {\em Kernel estimation in high-energy physics}, Comput. Phys. Commun. {\bf 136} (2001) 198.


\bibitem{combine}G. D'Agostini, {\em On the use of the covariance matrix to fit correlated data},  Nucl. Instrum. Methods Phys. Res., Sect. A {\bf 346} (1994) 306. 

\bibitem{PIDBelle2} Belle II Collaboration, {\em The Belle II Physics Book}, Prog. Theor. Exp. Phys. {\bf 2019} (2019) 123C01; {\bf 2020} (2020) 029201(E).

\bibitem{roohistpdf} I. Antcheva {\it et al.}, {\em ROOT — A C++ framework for petabyte data storage, statistical analysis and visualization},  Comput. Phys. Commun. {\bf 180} (2009) 2499.



\end{thebibliography}
\end{document}